\documentstyle[psfig]{mn}

\newif\ifAMStwofonts

\ifoldfss
  \newcommand{\rmn}[1] {{\rm #1}}

  \ifCUPmtlplainloaded \else
    \NewTextAlphabet{textbfit} {cmbxti10} {}
    \NewTextAlphabet{textbfss} {cmssbx10} {}
    \NewMathAlphabet{mathbfit} {cmbxti10} {} 
    \NewMathAlphabet{mathbfss} {cmssbx10} {} 
  \fi
  \ifAMStwofonts
    \ifCUPmtlplainloaded \else
      \NewSymbolFont{upmath} {eurm10}
      \NewSymbolFont{AMSa} {msam10}
      \NewMathSymbol{\upi}     {0}{upmath}{19}
      \NewMathSymbol{\umu}     {0}{upmath}{16}
      \NewMathSymbol{\upartial}{0}{upmath}{40}
      \NewMathSymbol{\leqslant}{3}{AMSa}{36}
      \NewMathSymbol{\geqslant}{3}{AMSa}{3E}

      \let\leq=\leqslant 
       
    \fi
  \fi
\fi 

\ifnfssone
  \newmathalphabet{\mathit}
  \addtoversion{normal}{\mathit}{cmr}{m}{it}
  \addtoversion{bold}{\mathit}{cmr}{bx}{it}
  \newcommand{\rmn}[1] {\mathrm{#1}}

  \newmathalphabet{\mathbfit} 
  \addtoversion{normal}{\mathbfit}{cmr}{bx}{it}
  \addtoversion{bold}{\mathbfit}{cmr}{bx}{it}
  \newmathalphabet{\mathbfss} 
  \addtoversion{normal}{\mathbfss}{cmss}{bx}{n}
  \addtoversion{bold}{\mathbfss}{cmss}{bx}{n}
  \ifAMStwofonts
    \ifCUPmtlplainloaded \else
      \UseAMStwoboldmath
      \makeatletter
      \new@mathgroup\upmath@group
      \define@mathgroup\mv@normal\upmath@group{eur}{m}{n}
      \define@mathgroup\mv@bold\upmath@group{eur}{b}{n}
      \edef\UPM{\hexnumber\upmath@group}
      \new@mathgroup\amsa@group
      \define@mathgroup\mv@normal\amsa@group{msa}{m}{n}
      \define@mathgroup\mv@bold\amsa@group{msa}{m}{n}
      \edef\AMSa{\hexnumber\amsa@group}
      \makeatother
      \mathchardef\upi="0\UPM19
      \mathchardef\umu="0\UPM16
      \mathchardef\upartial="0\UPM40
      \mathchardef\leqslant="3\AMSa36
      \mathchardef\geqslant="3\AMSa3E

      \let\leq=\leqslant 

    \fi
  \fi
\fi 

\ifnfsstwo
  \newcommand{\rmn}[1] {\mathrm{#1}}

  \DeclareMathAlphabet{\mathbfit}{OT1}{cmr}{bx}{it}
  \SetMathAlphabet\mathbfit{bold}{OT1}{cmr}{bx}{it}
  \DeclareMathAlphabet{\mathbfss}{OT1}{cmss}{bx}{n}
  \SetMathAlphabet\mathbfss{bold}{OT1}{cmss}{bx}{n}
  \ifAMStwofonts
    \ifCUPmtlplainloaded \else
      \DeclareSymbolFont{UPM}{U}{eur}{m}{n}
      \SetSymbolFont{UPM}{bold}{U}{eur}{b}{n}
      \DeclareSymbolFont{AMSa}{U}{msa}{m}{n}
      \DeclareMathSymbol{\upi}{0}{UPM}{"19}
      \DeclareMathSymbol{\umu}{0}{UPM}{"16}
      \DeclareMathSymbol{\upartial}{0}{UPM}{"40}
      \DeclareMathSymbol{\leqslant}{3}{AMSa}{"36}
      \DeclareMathSymbol{\geqslant}{3}{AMSa}{"3E}

      \let\leq=\leqslant 

    \fi
  \fi
\fi 

\ifCUPmtlplainloaded \else
  \ifAMStwofonts \else 
    \def\upi{\pi}
    \def\umu{\mu}
    \def\upartial{\partial}
  \fi
\fi

\title{Physical conditions in QSO absorbers from fine-structure absorption lines}
\author[Silva \& Viegas]
       {A. I. Silva \thanks{E-mail: ignacioalex@yahoo.com .}
	and S. M. Viegas \thanks{E-mail: viegas@iagusp.usp.br .}\\
        Instituto Astron\^omico e Geof\'\i sico, Universidade de S\~ao Paulo,
	Av. Miguel St\'efano, 4200, 04301-904 S\~ao Paulo SP, Brazil}
\date{Accepted .
      Received }

\pagerange{\pageref{firstpage}--\pageref{lastpage}}
\pubyear{2000}

\begin{document}

\maketitle

\label{firstpage}

\begin{abstract}
We calculate theoretical population ratios of the ground fine-structure
levels of some atoms/ions which typically exhibit
UV lines in the spectra of QSO absorbers redward the Ly-$\alpha$ forest:
C$^0$, C$^+$, O$^0$, Si$^+$ and Fe$^+$.
The most reliable atomic data available is employed and a variety of excitation
mechanisms considered: collisions with several particles in the medium,
direct excitation by photons from the cosmic microwave background radiation
(CMBR) and fluorescence induced by a UV field present.

The theoretical population ratios are confronted with the corresponding
column density ratios of \hbox{C\,{\sc i}} and \hbox{C\,{\sc ii}} lines
observed in damped Ly-$\alpha$ (DLA) and Lyman Limit (LL) systems
collected in the
recent literature to infer their physical conditions.

The volumetric density of neutral hydrogen in DLA systems is constrained
to be lower than tens of cm$^{-3}$ (or a few cm$^{-3}$ in the best cases)
and the UV radiation field intensity must be lower than two orders of
magnitude the radiation field of the Galaxy (one order of magnitude in the
best cases). Their characteristic sizes are higher than a few pc (tens of
pc in the best cases) and lower limits for their total masses vary from
$10^0$ to $10^5$ solar masses.

For the only LL system in our sample, the electronic density is constrained
to be $n_e<0.15$ cm$^{-3}$. We suggest that the fine-structure lines
may be used to discriminate between the current accepted picture of the
UV extragalatic background as the source of ionization in these systems
against a local origin for the ionizing radiation as supported by some
authors.

We also investigate the validity of the temperature-redshift relation
of the CMBR predicted by the standard model and study the case for
alternative models.
\end{abstract}

\begin{keywords}
quasars: absorption lines - cosmic microwave background - atomic processes.
\end{keywords}

\section{Introduction}

Typically in the spectra of bright QSOs, several lines may be identified
as beeing due to absorption of intervening material situated along the line of sight.
These lines originate when the continuum radiation meets a common atom or ion in its
ground state that partially absorbs it, leaving its imprint in the emiting QSO's spectrum.
However, if some excitation mechanism is present in the absorbing region, then a small
fraction of atoms or ions will also be found populated in their lowest-lying excited levels.
Therefore, in addition to absorption lines arising from the atom/ion's ground state, one
may also expect to detect weaker lines arising from excited levels.

It has been long pointed out that fine-structure absorption lines arising from the ground
and lowest-lying excited energy levels of common atoms/ions may be used as an
indicator of the physical conditions in the gas \cite{BW,SP}.

If we model the absorbing region as a single, homogeneous cloud, then the ratio of the
volumetric densities of atoms/ions populated in excited states $n^*$ to atoms/ions
in the ground state $n$ will match the corresponding column density ratios:

\begin{equation}
\frac{\,\,n^*}{n} = \frac{\,\,\,N^*}{N} \ .
\label{eq:ratio}
\end{equation}

For example, the column densities of C$^+$ ions populated in their ground
$^2\mathrm{P}^o_{1/2}$ and first excited $^2\mathrm{P}^o_{3/2}$ levels may be inferred
from the equivalent widths of the corresponding
2s$^2$2p $^2\mathrm{P}^o_{1/2}\rightarrow$ 2s2p$^2$ ${^2\mathrm{D}^e_{3/2}}$ and
2s$^2$2p $^2\mathrm{P}^o_{3/2}\rightarrow$ 2s2p$^2$ ${^2\mathrm{D}^e_{5/2}}$ UV lines
at 1334.5 \AA\ and 1335.7 \AA, respectively.

The lefthand side of equation (\ref{eq:ratio}) in turn, may be evaluated theoretically as a
function of the physical conditions in the medium by solving the detailed
equations of statistical equilibrium. It will in general depend on the intensities of
several competing excitation mechanisms, such as spontaneous decay, collisions
with particles present in the medium or induced by radiation
(the later either directly or by fluorescence).

The effectiveness of using column density ratios deduced from fine-structure lines to infer the
basic parameters of a given excitation mechanism will depend on its relative importance
to other processes contributing to the excitation of the fine structure levels.
If collisions by a given particle
dominate, one may expect to be able to infer its volumetric density; if
fluorescence dominates, one is capable of measuring the intensity of the radiation field
present, whereas if the dominating mechanism is direct excitation by photons of the
Cosmic Microwave Background Radiation (CMBR) one could measure its temperature.
This dominance of some process over another is determined not only by their relative
intensities, but also by an interplay of atomic physics input parameters.

The goal of this paper is to calculate theoretical population ratios of fine-structure
levels of atoms/ions commonly found in QSO spectra, and to use them to estimate the
physical conditions in QSO absorbers.

In section \ref{section:populationratios} we describe how the equations of statistical
equilibrium were solved and details on the calculations for each selected atom/ion,
namely: C$^0$, C$^+$, Si$^+$, O$^0$ and Fe$^+$. In section \ref{section:physicalconditions}
we gather recent column density ratios data taken from the literature and use the results
obtained in the previous section to determine the physical conditions in Damped Lyman-$\alpha$
(DLA) and Lyman Limit (LL) QSO absorption line systems.
For the DLA systems we also derive their characteristic sizes and masses.
The main conclusions are sketched in section \ref{section:conclusions}.

\section{Atomic Physics}
\label{section:populationratios}

In this section we calculate the population ratios of fine-structure levels for
five atoms/ions of interest: C$^0$, C$^+$, O$^0$, Si$^+$ and Fe$^+$.
The first two already have their fine-structure lines observed with currently available
high resolution spectrographs (see section \ref{section:conclusions} below).
The other atoms/ions are typically observed in QSO absorption spectra, since they have
resonant lines longward the Ly-$\alpha$ line at 1216 \AA\ (so that they will not always
fall into the Ly-$\alpha$ forest region of the spectrum). They also have their
ground term split into fine-structure levels, and once the
lines arising from excited levels are detected in future generations of more powerful
telescopes, they may also provide useful physical condition information.

Let us now briefly outline the basic procedures to calculate the fine-structure levels
population ratios, and next discuss each particular atom/ion in greater detail.

\subsection{The statistical equilibrium equations}

In order to calculate the level populations of a given
atom/ion, we make two basic assumptions :

\begin{enumerate}

\item The rates of processes involving ionization stages other than the atom/ion
being considered (such as direct photoionization or recombination, charge exchange reactions,
collisional ionization, etc.) are slow compared to bound-bound rates.

\item All transitions considered are optically thin.

\end{enumerate}

In steady state regime the sum over all processes populating a given level $i$ will
be balanced by the sum over all processes depopulating it. Assuming that the two
conditions listed above are met, this can be written:

\begin{equation}
\sum_j n_j Q_{ji} = n_i \sum_j Q_{ij} \ ,
\label{eq:sum}
\end{equation}

where $n_i$ is the volume density of atoms or ions in level $i$.
We have defined the total rates taking the atom/ion from level $i$ to level $j$ as:

\begin{equation}
Q_{ij} \equiv A_{ij} + B_{ij}u_{ij} + \Gamma_{ij} + \sum_k n^k q_{ij}^k \ ,
\end{equation}

where the coefficients $A_{ij}$ are transition probabilities, $B_{ij}$ are Einstein
coefficients, $u_{ij}$ are the energy densities of the radiation field at the frequency
of the transition $\nu_{ij}$, $\Gamma_{ij}$ are indirect excitation rates by fluorescence,
$n^k$ are the volumetric densities of a given collision particle
(usually $k = e^-, p^+, H^0, {He}^0, H_2, \dots$ , depending whether the medium
is primarily ionized or neutral) and $q_{ij}^k$ are the collision rates by some particle $k$.
We have set $A_{ij}=0$ for $i\leq j$ and
$B_{ii}=u_{ii}=\Gamma_{ii}=q_{ii}^k=0$.

Hereafter we shall abbreviate:

\begin{equation}
K_{ij} \equiv B_{ij} u_{ij} \ .
\end{equation}

The indirect excitation rates are defined as \cite{SV2000}:

\begin{equation}
\Gamma_{ij} \equiv \sum_{\mu} K_{i\mu} \frac {A_{\mu j}+K_{\mu j}}
{\sum_{g=1}^m \left( A_{\mu g} + K_{\mu g} \right)} \ ,
\end{equation}

i.e., we have the situation in which the atom/ion - in one of its $m$
lowest energy levels, $i$ - is photoexcited to some higher energy level $\mu$ and
next decays - either spontaneously or by stimulated emission - back to some
other level $j$ among the lowest $m$. The sum extends over all possible upper levels.

The fine-structure levels may also be directly populated by the CMBR. In that case one
must add to the energy densities $u_{ij}$ the contribution from a black body radiation
field redshited to a temperature (see, for instance, Kolb \& Turner 1990):

\begin{equation}
T=T_0 \left( 1+z \right) \ ,
\label{eq:Tlaw}
\end{equation} 

where $T_0=2.725\pm 0.001$ K ($1\sigma$ error) is the current value of the CMBR temperature
as determined from the {\it COBE FIRAS \/} instrument \cite{Mather99,Smoot}.

We caution, however, that this relation remains yet observationally unprooven.
In section \ref{section:CMBR} we review the currently available pieces of evidence.

Equation (\ref{eq:sum}) is the system of statistical equilibrium equations that
must be solved in order to compute the population ratios of the fine-structure levels.
If we model the atom/ion as beeing composed of $m$ levels, then we must deal with a
system of $m-1$ equations.

In order to numerically solve this system we have built
a Fortran 90 code - {\sc popratio} \cite{SV2000}- that reads in the basic atomic physics parameters
and automatically computes the rates for all the processes beeing considered.
The code is very flexible, allowing the user to account
for an arbitrary number of levels and processes.

The code, as well as the input files for the atoms/ions considered in this paper, are
available upon request from one of the authors
(AIS)\footnote{It is also available at the following http URL:
http://www.iagusp.usp.br/\~\,alexsilv/popratio\,.}.
It may also be used in other
astronomical applications, such as calculating intensity ratios of collisionally
excited emission lines (such as coronal emission lines) and computing cooling
rates due to collisional excitation.

Next, we describe the computations for each atom/ion considered in greater detail.
As the population ratios of the fine-structure levels will be strongly dependent
upon several atomic physics parameters, it is essential to search the literature for
the most up to date values. In this work, we give precedence to results obtained
recently by two large international collaborations: the Opacity Project \cite{Opacity}
and the Iron Project \cite{Iron}.

For reasons of space, we illustrate the results obtained for the population ratios
of the fine-structure levels under a limited range of physical conditions only.
We urge the user to make use of the numerical code in order to get accurate predictions
in his/her applications.

\subsection{The atom C$^0$}
\label{C0}

The ground state of the C$^0$ atom is comprised of the 2s$^2$2p$^2$ $^3\mathrm{P}^e_{0,1,2}$
triplet levels. The energies of the fine-structure excited levels relatively to the ground state
are 16.40 cm$^{-1}$ and 43.40 cm$^{-1}$. The transition probabilities are
$A_{10}=7.932\ 10^{-8}\ \mathrm{s}^{-1}$,
$A_{20}=2.054\ 10^{-14}\ \mathrm{s}^{-1}$ and
$A_{21}=2.654\ 10^{-7}\ \mathrm{s}^{-1}$.

Our model atom includes the five lowest energy levels:
2s$^2$2p$^2$ $^3\mathrm{P}^e_{0,1,2}$,
2s$^2$2p$^2$ $^1\mathrm{D}^e_2$ and
2s$^2$2p$^2$ $^1\mathrm{S}^e_0$.
The energies were taken from
Moore \shortcite{MooreCI} and the transition probabilities from the Iron Project calculation
of Galav\'\i s, Mendoza and Zeippen \shortcite{AijCI}.

The CMBR will be an important excitation mechanism for the first excited
$^3\mathrm{P}^e_1$ level, since it is so closely separated from the ground level.
Assuming the temperature-redshift relation as given by equation (\ref{eq:Tlaw}), the CMBR
spectrum will peak at the first-excited level frequency at a redshift $z\sim 2$.
Table \ref{table:Kij} gives the excitation rates of the C$^0$ fine-structure levels
as a function of redshift, again assuming the temperature-redshift relation
(\ref{eq:Tlaw}).

\begin{table}
\caption{Excitation rates $K_{JJ'}$ of the C$^0$, C$^+$ and O$^0$
fine-structure levels by the CMBR.
We have assumed the temperature-redshift relation as predicted by the standard model
(see text).}
\label{table:Kij}
\begin{tabular}{@{}ccccc}
  & \multicolumn{2}{c}{C$^0$} & C$^+$ & O$^0$ \\
z & $K_{01}$ (s$^{-1}$) & $K_{02}$ (s$^{-1}$) & $K_{\frac{1}{2}\frac{3}{2}}$ (s$^{-1}$) 
& $K_{21}$ (s$^{-1}$) \\
\hline
0 & 4.2\ 10$^{-11}$ & 1.2\ 10$^{-23}$ & 1.4\ 10$^{-20}$ & 3.0\ 10$^{-41}$ \\
1 & 3.2\ 10$^{-9}$  & 1.1\ 10$^{-18}$ & 2.5\ 10$^{-13}$ & 4.0\ 10$^{-23}$ \\
2 & 1.4\ 10$^{-8}$  & 5.0\ 10$^{-17}$ & 6.6\ 10$^{-11}$ & 4.4\ 10$^{-17}$ \\
3 & 3.1\ 10$^{-8}$  & 3.4\ 10$^{-16}$ & 1.1\ 10$^{-9}$ & 4.6\ 10$^{-14}$ \\
4 & 5.1\ 10$^{-8}$  & 1.1\ 10$^{-15}$ & 5.7\ 10$^{-9}$ & 3.0\ 10$^{-12}$ \\
5 & 7.4\ 10$^{-8}$  & 2.3\ 10$^{-15}$ & 1.7\ 10$^{-8}$ & 4.8\ 10$^{-11}$
\end{tabular}
\end{table}

\begin{figure}
\vbox{
\psfig{figure=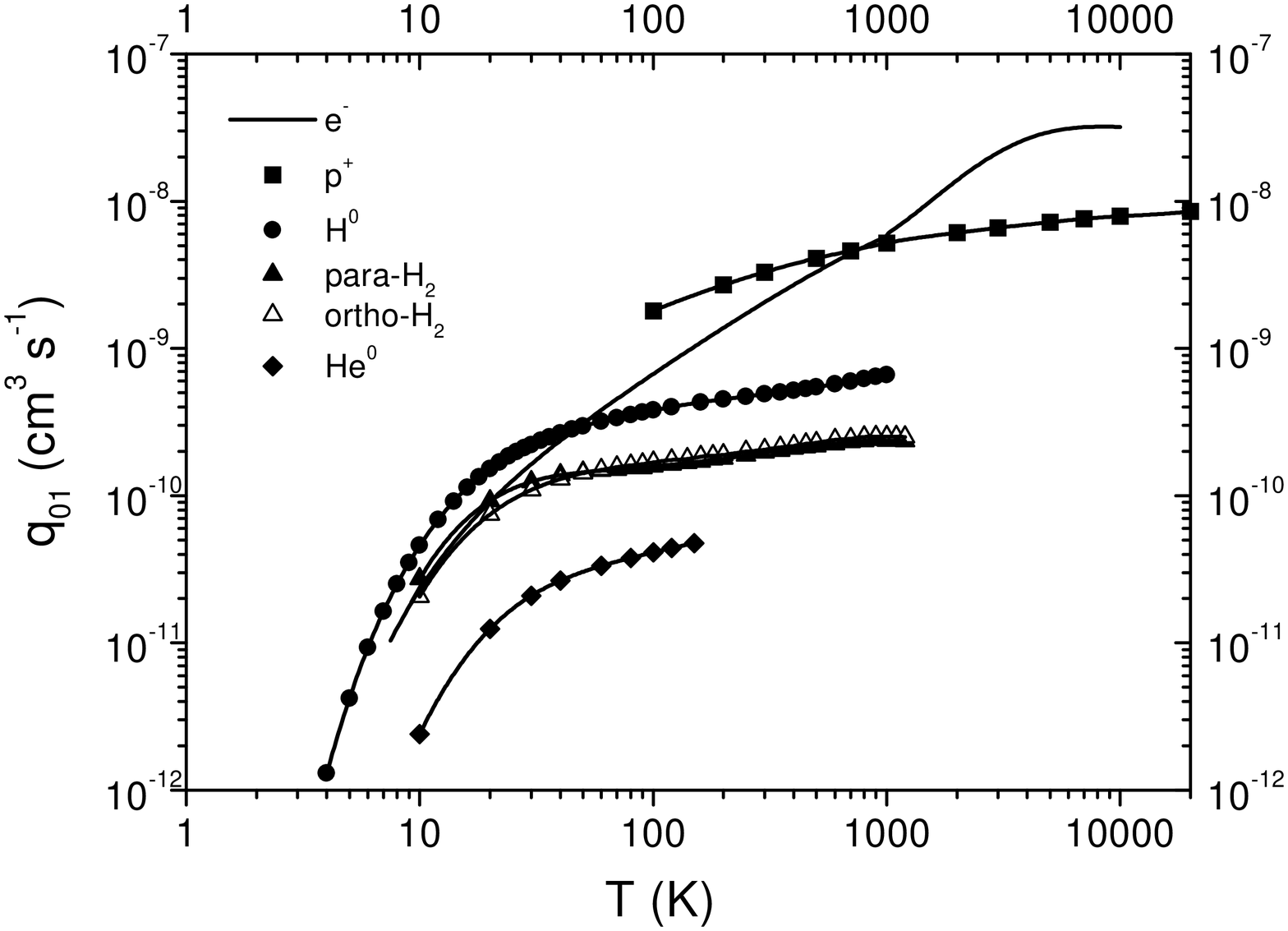,width=9.5cm}
\psfig{figure=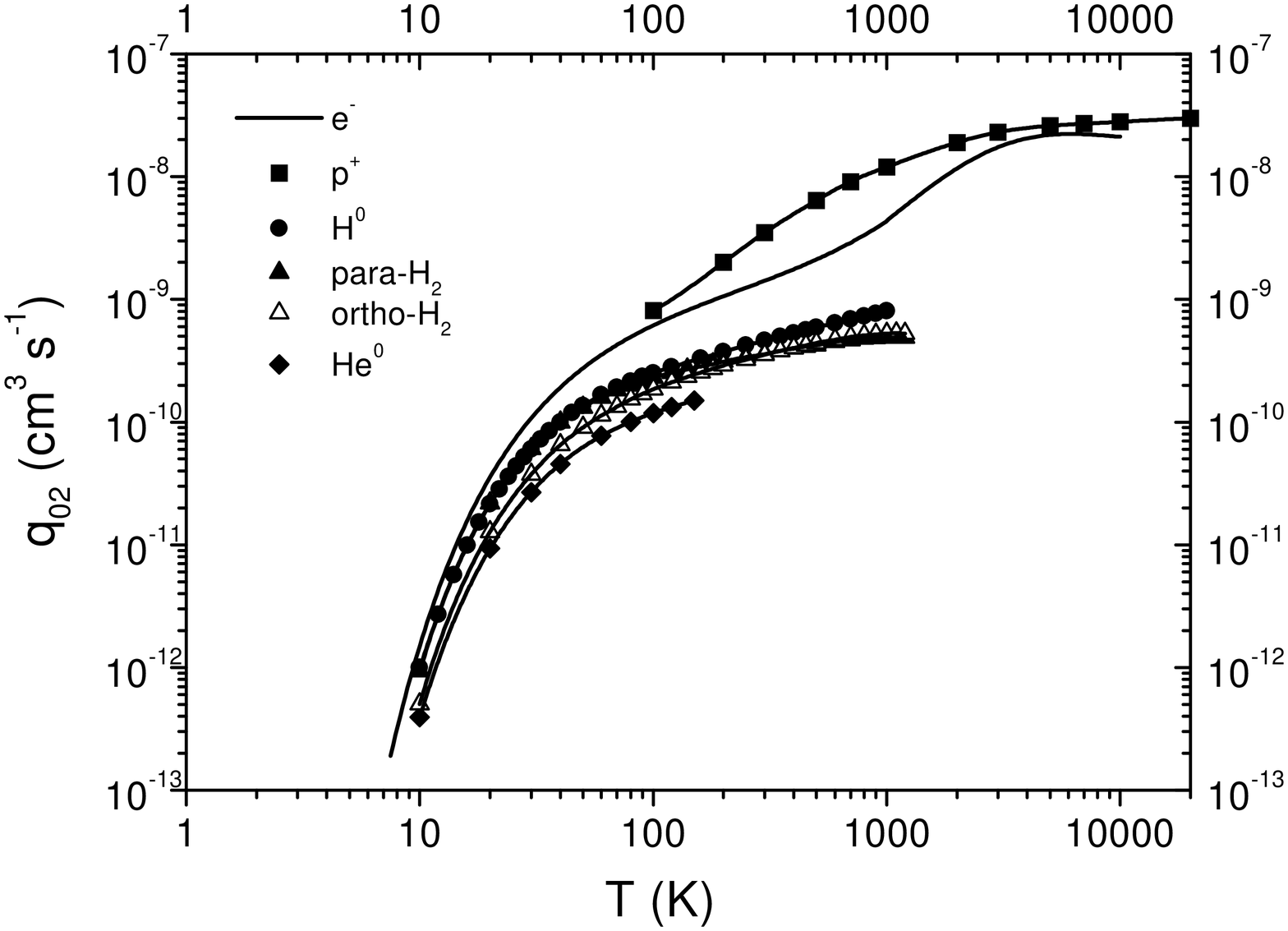,width=9.5cm}
\psfig{figure=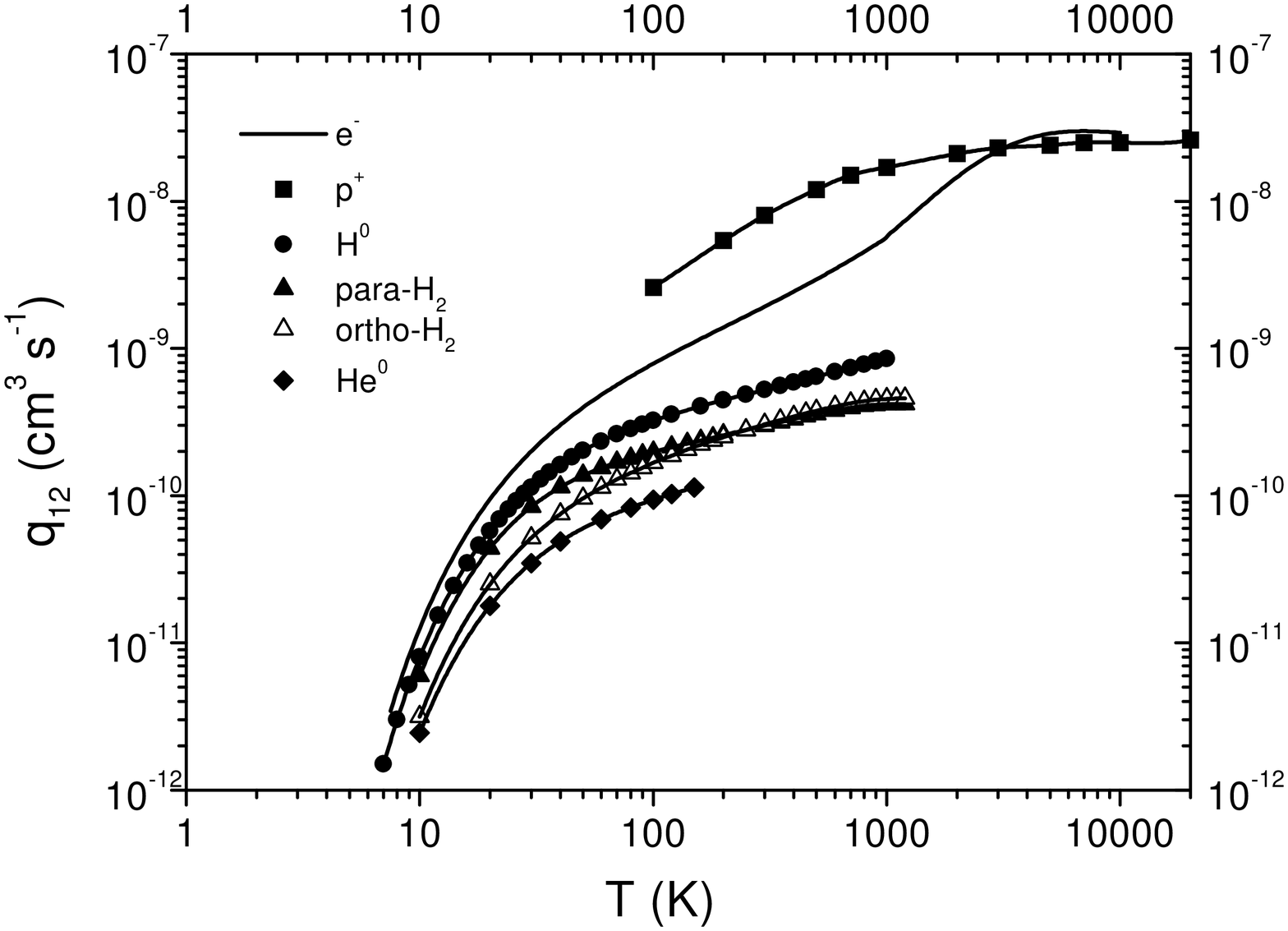,width=9.5cm}
}
\caption{Excitation rates $q_{JJ'}=q({^3P^e_J}\rightarrow{^3P^e_{J'}})$
of the C$^0$ fine-structure levels by collisions with various particles.
The points - taken from the literature cited in the text - are interpolated by
cubic splines.}
\label{figure:qijCI}
\end{figure}

The fine-structure transitions may also be induced by collisions.
Fig. \ref{figure:qijCI} shows the collision rates for the most important collision particles.
The rates for collisions by protons were taken from Roueff \& Le Bourlot
\shortcite{q_proton_CI}, by neutral hydrogen from Launay \& Roueff \shortcite{q_H0_CI},
by molecular hydrogen from Schr\"oder et al. \shortcite{q_H2_CI} and by neutral
helium from Staemmler \& Flower \shortcite{q_He_CI}. For the rates by collisions with
electrons we have employed the analytic fits given by Johnson, Burke \& Kingston
\shortcite{q_electron_CI}. We point out that the similar plot in Roueff \& Le Bourlot's
paper comparing collision rates by protons and electrons is incorrect, since an error
has crept in their figure, and they compare excitation and de-excitation rates
(Roueff, private comm.).

We have also considered the effect of excitation of the upper $^1\mathrm{D}^e_2$ and
$^1\mathrm{S}^e_0$ levels by collisions with electrons.
We took the analytic fits to the Maxwellian-averaged collision strengths $\gamma$ for transitions
involving these levels given by P\'equignot \& Aldrovandi \shortcite{high_e_CI}, and
transformed them from LS coupling to the fine-structure levels according to their
statistical weights:

\begin{eqnarray}
\label{eq:LSindividual}
\gamma \left( {^3\mathrm{P}^e_J} \rightarrow {^1\mathrm{D}^e_2} \right) & = &
\frac{2J+1}{9} \gamma \left( {^3\mathrm{P}^e} \rightarrow {^1\mathrm{D}^e} \right) \\
\gamma \left( {^3\mathrm{P}^e_J} \rightarrow {^1\mathrm{S}^e_0} \right) & = &
\frac{2J+1}{9} \gamma \left( {^3\mathrm{P}^e} \rightarrow {^1\mathrm{S}^e} \right)
\nonumber \ .
\end{eqnarray}

However, the inclusion of these levels can hardly influence the population of the
$^3\mathrm{P}^e$ fine-structure levels at temperatures prevailing in ionization
regions where the atom C$^0$ is likely to be found. For example, even for temperatures
as high as $T=10^4\ K$ the population ratio of the $^3\mathrm{P}^e_1$ level relatively to
the ground state will increase by no more than 5 percent (10 percent for the
$^3\mathrm{P}^e_2$ level). The test calculations were done taking into account only
collisions by electrons (and spontaneous decays);
if this is not the main excitation mechanism, then the error
will be significantly smaller.

Excitation of the fine-structure levels by fluorescence was also investigated.
We consider 108 allowed UV transitions involving the ground $^3\mathrm{P}^e$
levels and upper levels listed in the compilation of Verner, Verner \& Ferland
\shortcite{Verner}, which is based on Opacity Project calculations.
If we adopt the radiation field of the Galaxy \cite{Gondhalekar}, then the corresponding
indirect excitation rates will be $\Gamma_{01}=3.5\ 10^{-10}\ \mathrm{s}^{-1}$
and $\Gamma_{02}=2.8\ 10^{-10}\ \mathrm{s}^{-1}$.

\begin{figure*}
\vbox{
\hbox{
\psfig{figure=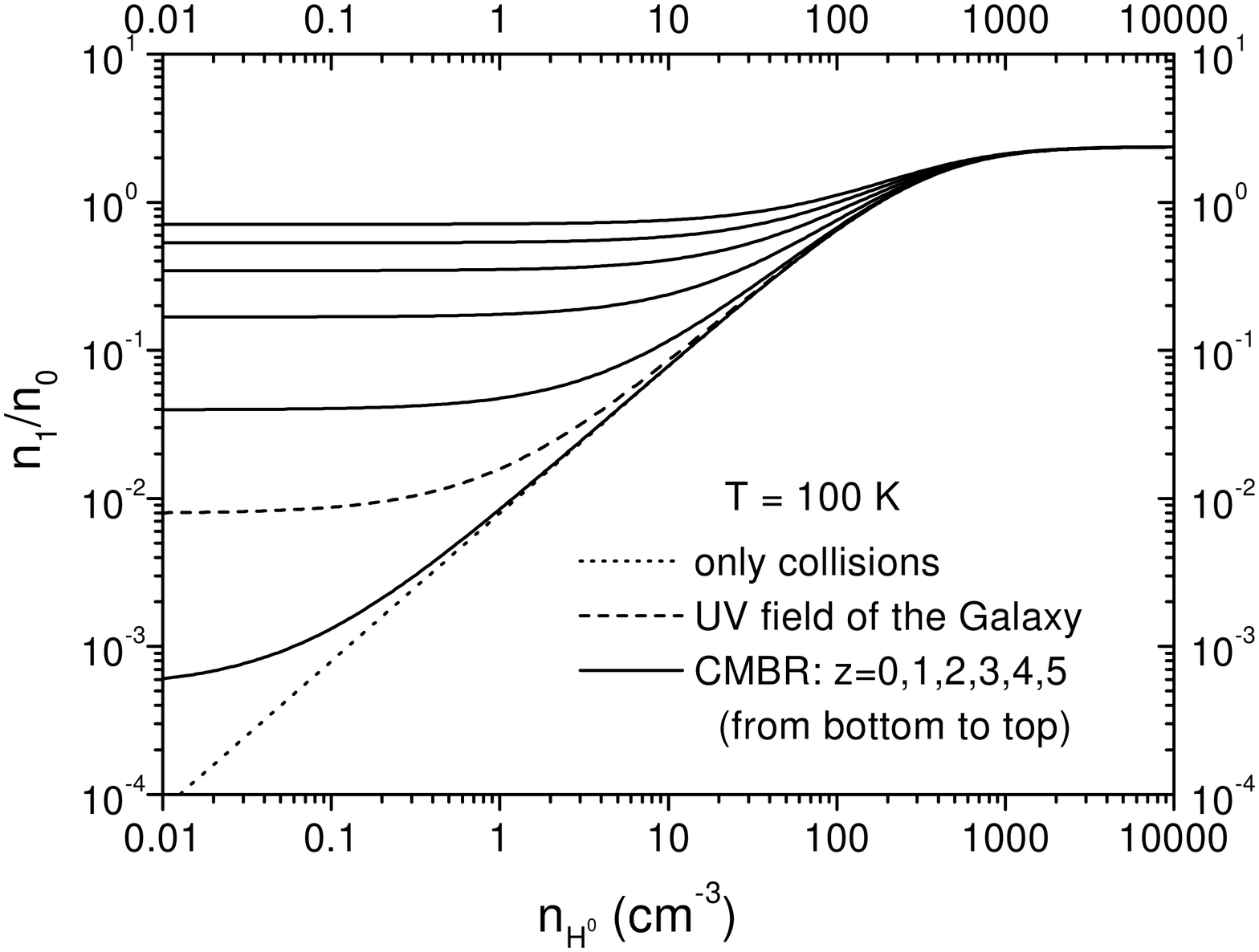,width=9.5cm}
\psfig{figure=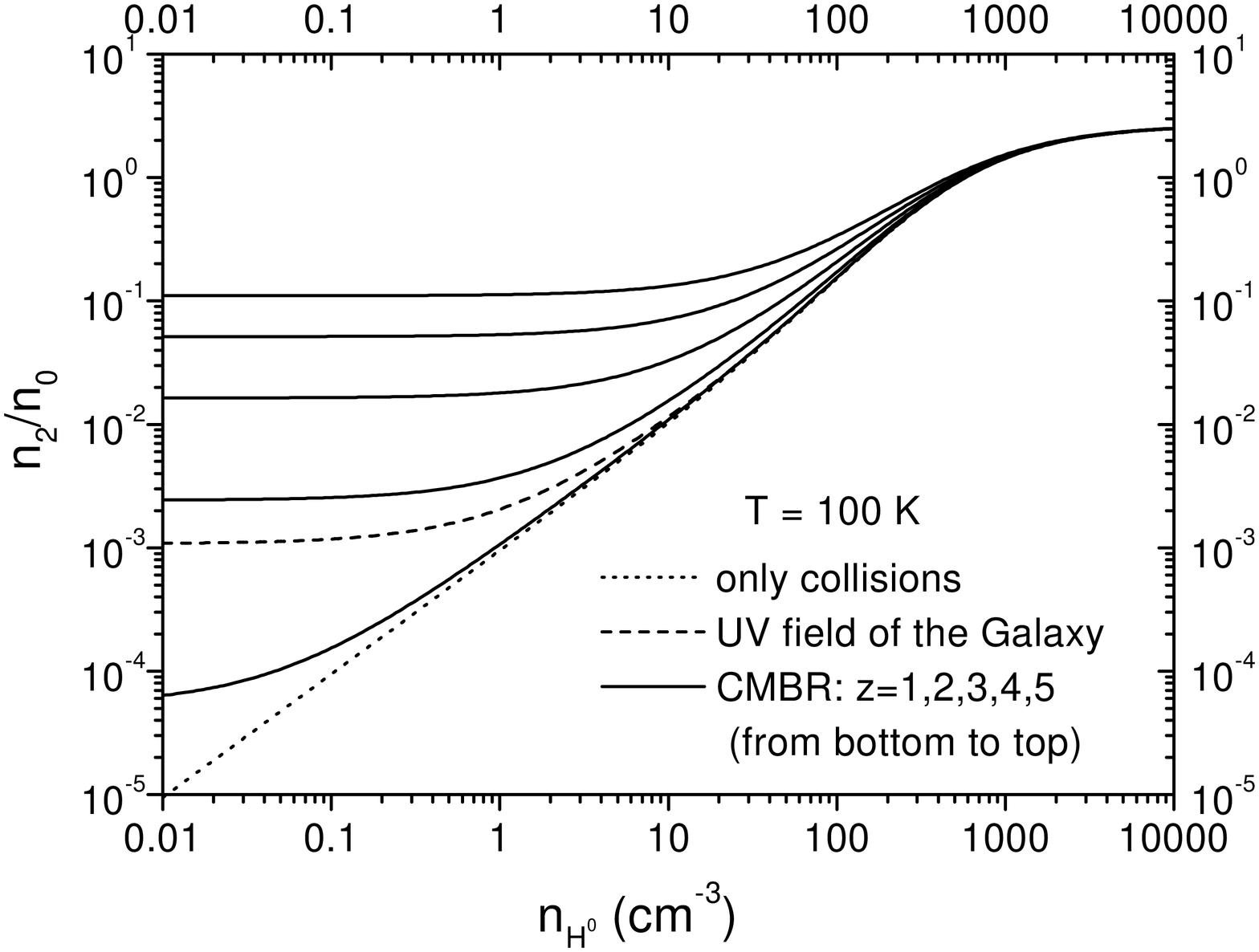,width=9.5cm}
}
\hbox{
\psfig{figure=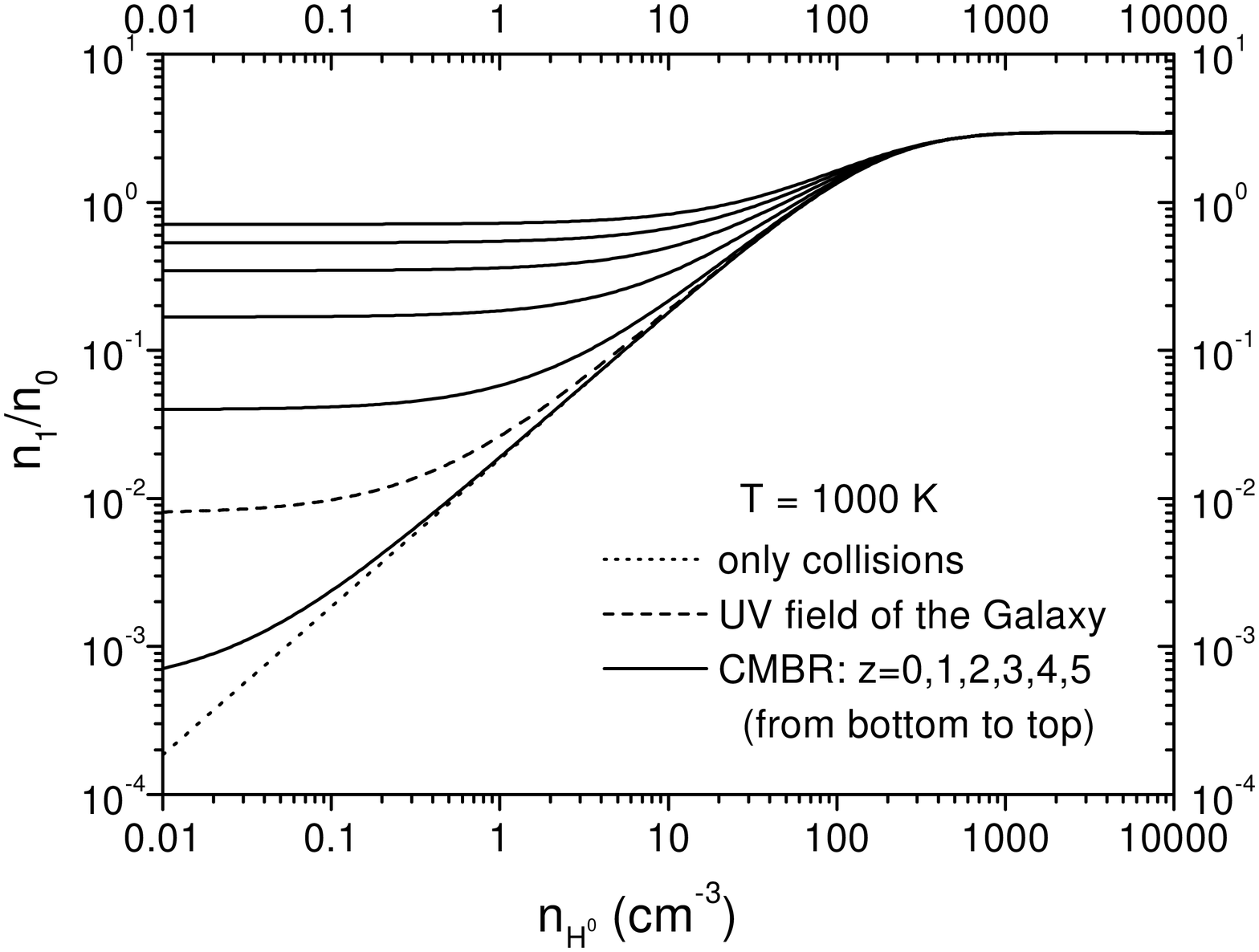,width=9.5cm}
\psfig{figure=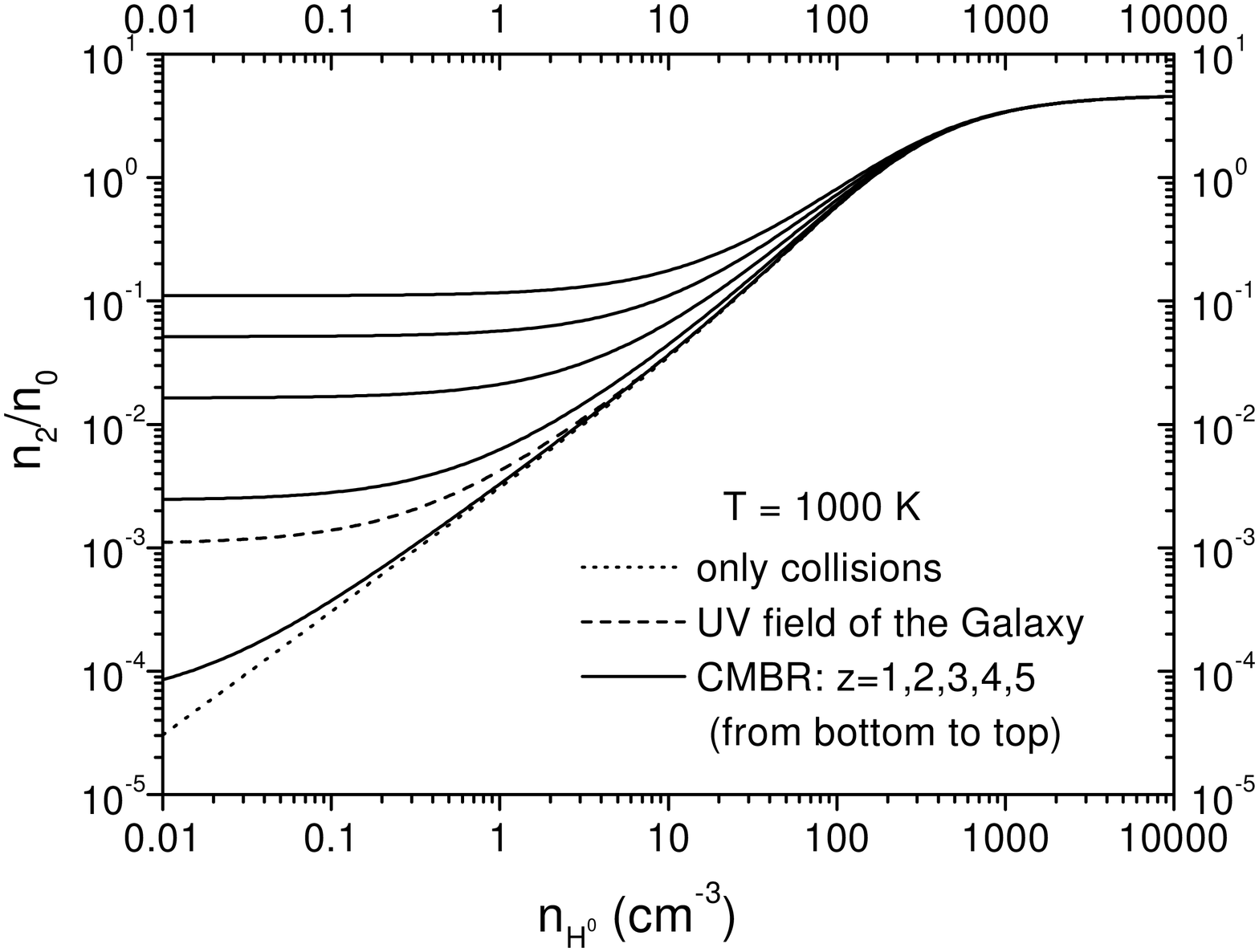,width=9.5cm}
}
}
\caption{Population ratios of the C$^0$ fine-structure levels relatively to the
ground state $n_J/n_0 = n({^3\mathrm{P}^e_J})/n({^3\mathrm{P}^e_0})$
calculated under various physical conditions. The curve for the $n_2/n_0$ population ratio
for $z=0$ coincides with the curve taking only collisions into account.}
\label{figure:popCI}
\end{figure*}

In fig. \ref{figure:popCI} we have plotted the population ratios of the C$^0$ fine-structure
levels taking into account collisions by hydrogen atoms (the main collision partner
in ionization regions where the atom C$^0$ is found), the CMBR and fluorescence
induced by the radiation field of the Galaxy.

We compared our results with the previous calculations by Keenan \shortcite{KeenanCI}.
He considered the effect of collisions by electrons and hydrogend atoms, as well as
fluorescence induced by the radiation field of the Galaxy. Test calculations revealed
good overall agreement with the values obtained by Keenan, with differences typically less
than 15 percent.

\subsection{The ion C$^+$}

The ground state of the C$^+$ ion consists of the
2s$^2$2p $^2\mathrm{P}^o_{\frac{1}{2},\frac{3}{2}}$
doublet levels. The energy of the fine-structure excited level relatively to the ground state
is  63.42 cm$^{-1}$, and the transition probability is
$A_{\frac{3}{2}\frac{1}{2}}=2.291\ 10^{-6}\ \mathrm{s}^{-1}$.

Our model ion includes the five lowest LS terms:
2s$^2$2p $^2\mathrm{P}^o$
and the 2s 2p$^2$ configurations
$^4\mathrm{P}^e$,
$^2\mathrm{D}^e$,
$^2\mathrm{S}^e$ and
$^2\mathrm{P}^e$,
making a total of ten levels when the fine-structure splitting is accounted for. 
The energies were taken from Moore \shortcite{MooreCI} and the transition probabilities
from the Iron Project calculation of Galav\'\i s, Mendoza \& Zeippen \shortcite{AijCII}.

As the fine-structure levels of C$^+$ are more separated than the C$^0$ levels,
the CMBR will play a significant role at higher redshifts only, as one can see
from the excitation rates given in table \ref{table:Kij} \footnote{Hereafter we
shall assume as a working hypothesis the temperature-redshift relation
predicted by the standard model.}.

\begin{figure}
\psfig{figure=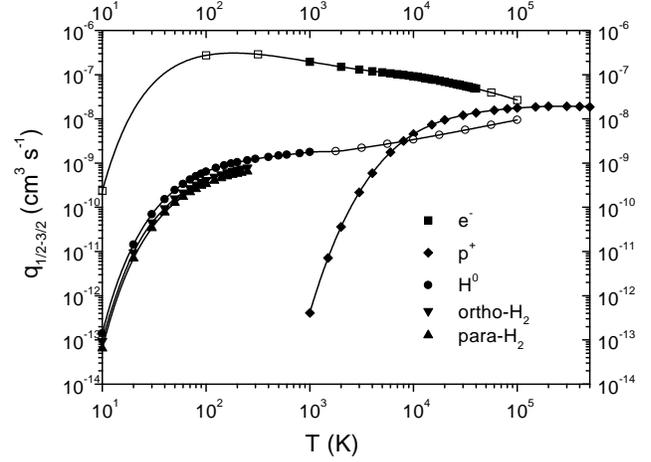,width=9.5cm}
\caption{Excitation rates $q_{\frac{1}{2}\frac{3}{2}}=
q({^2P^o_{\frac{1}{2}}}\rightarrow{^2P^o_{\frac{3}{2}}})$
of the C$^+$ fine-structure level by collisions with various particles.
The points - taken from the literature cited in the text - are interpolated by
cubic splines.}
\label{figure:qijCII}
\end{figure}

We take into account collisional excitation of the fine-structure levels with several
particles. For the Maxwellian-averaged collision strengths for collisions by electrons
we have adopted the calculation of Blum \& Pradhan \shortcite{gammaCII}.
As their results differ by only 2 percent from the earlier calculation of Keenan et al.
\shortcite{KeenanCII}, we have also included the later's results at temperature values not
covered by Blum \& Pradhan's calculation as a means of broadening the available
temperature range. We took excitation rates by collisions with hydrogen atoms from
Launay \& Roueff \shortcite{qH0_CII},
extrapolated to $T>1000$ K by Keenan et al. \shortcite{KeenanCII}.
Other collision particles taken into account are protons \cite{q_proton_CII} and
molecular hydrogen \cite{q_H2_CII}. Fig. \ref{figure:qijCII} compares the excitation
rates with the various particles.
 
We have complemented the work of Galav\'\i s et al. with
the allowed transitions listed in the compilation of Verner et al., making a total of 48
transitions involving the ground $^2$P$^o$ levels and upper levels. 
The indirect excitation rate by the UV radiation field of the Galaxy could then be
determined: $\Gamma_{\frac{1}{2}\frac{3}{2}}=9.3\ 10^{-11}\ \mathrm{s}^{-1}$.

In order to assess the relevance of the 2s 2p$^2$ configuration upper levels in the
relative population of the ground $^2\mathrm{P}^o_{\frac{1}{2},\frac{3}{2}}$ levels,
we have performed test calculations comparing our 10-level model ion with the
2-level ion. At high temperatures the 2s 2p$^2$ configuration levels
may be excited by collisions with hot electrons in the medium. However, the testcases
have shown that this effect does not contribute significantly to the excitation of the
$^2\mathrm{P}^o$ levels for temperatures $T\leq 30000$ K, where the discrepancies reach
about 5 percent.

Therefore, for temperatures lower than 30000 K, only two levels can be taken
into account. The population ratio of the excited fine-structure level relatively to
the ground level is then expressed by:

\begin{eqnarray}
\frac{n_{\frac{3}{2}}}{n_{\frac{1}{2}}} =
\frac{Q_{\frac{1}{2}\frac{3}{2}}}{Q_{\frac{3}{2}\frac{1}{2}}} & = &
\frac{K_{\frac{1}{2}\frac{3}{2}}+\Gamma_{\frac{1}{2}\frac{3}{2}}+
\sum_k n^k q^k_{\frac{1}{2}\frac{3}{2}}}
{A_{\frac{3}{2}\frac{1}{2}} + K_{\frac{3}{2}\frac{1}{2}}+\Gamma_{\frac{3}{2}\frac{1}{2}}+
\sum_k n^k q^k_{\frac{3}{2}\frac{1}{2}}} \\
& \cong & \frac{K_{\frac{1}{2}\frac{3}{2}}+\Gamma_{\frac{1}{2}\frac{3}{2}}+
\sum_k n^k q^k_{\frac{1}{2}\frac{3}{2}}}
{A_{\frac{3}{2}\frac{1}{2}} + \sum_k n^k q^k_{\frac{3}{2}\frac{1}{2}}} \nonumber \ .
\end{eqnarray}

The collisional de-excitation rates may be computed from the principle of detailed balance:

\begin{equation}
q_{\frac{3}{2}\frac{1}{2}} = \frac{1}{2} q_{\frac{1}{2}\frac{3}{2}} \mathrm{e}^{\frac{91.25}{T}} \ ,
\end{equation}

with $T$ expressed in K.

\begin{figure}
\vbox{
\psfig{figure=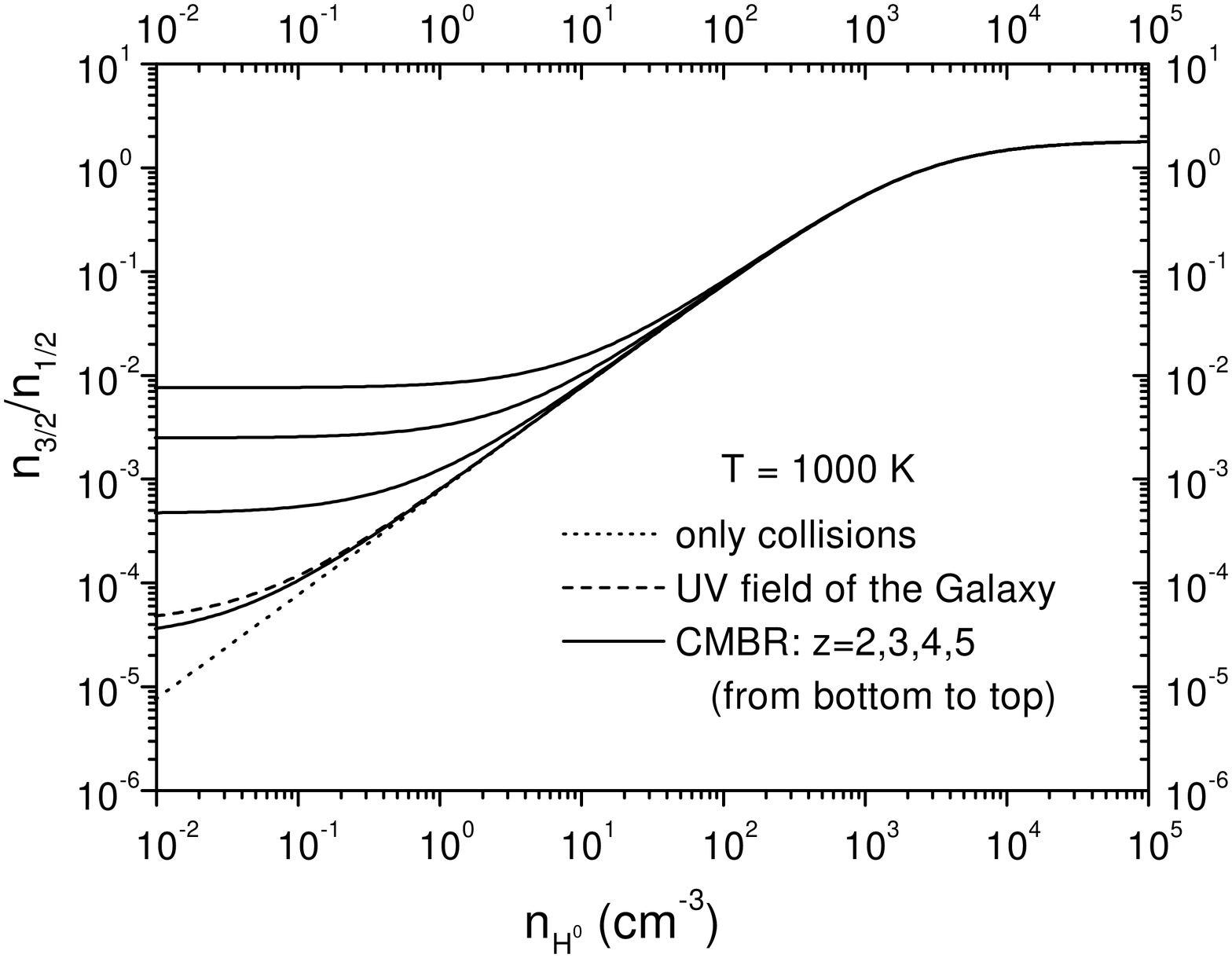,width=9.5cm}
\psfig{figure=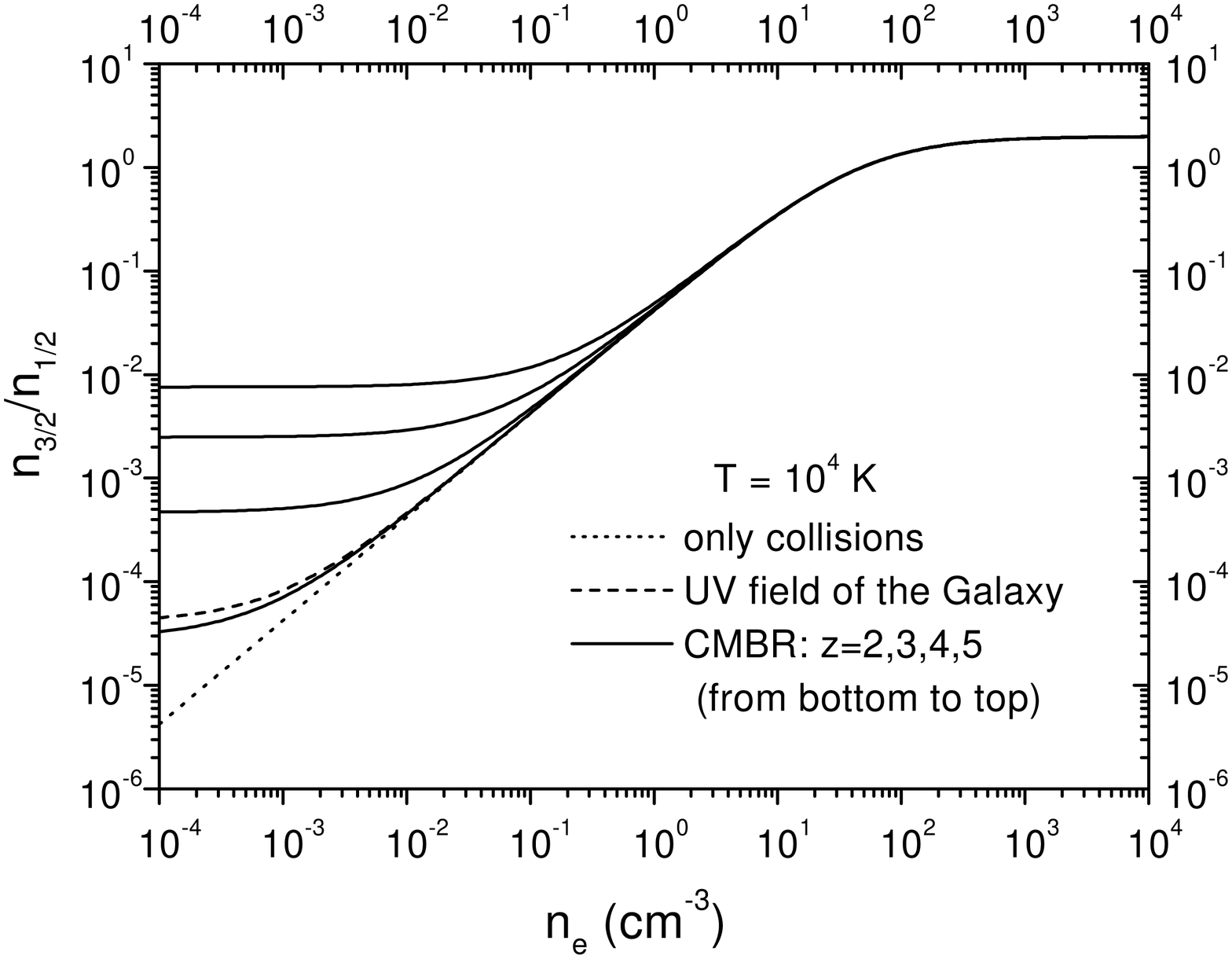,width=9.5cm}
}
\caption{Population ratio of the C$^+$ fine-structure level relatively to the
ground state $n_{\frac{3}{2}}/n_{\frac{1}{2}} =
n({^2\mathrm{P}^o_{\frac{3}{2}}})/n({^2\mathrm{P}^o_{\frac{1}{2}}})$
calculated under various physical conditions. The curves
for $z\leq 1$ coincide with the curve taking only collisions into account.
In the lower plot we have also taken proton collisions into account, $n_p=n_e$.}
\label{figure:popCII}
\end{figure}

In fig. \ref{figure:popCII} we have plotted the population ratio of the C$^+$
excited fine-structure level relatively to the ground state
under various physical conditions. As the ion C$^+$ may coexist in both
\hbox{H\,{\sc i}} and \hbox{H\,{\sc ii}} regions, we sample two cases of interest:
a neutral medium
at $T=1000$ K, and an ionized medium at $T=10000$ K. In the later case,
in addition to collisions by electrons, we also consider proton collisions
and set $n_p=n_e$.
However, at $T=10000$ K their effect on the relative population ratio
is only marginal (at the 5 percent level).

Previous work on the population of the C$^+$ fine-structure levels
taking into account fluorescence and collisions by electrons and
hydrogen atoms was accoplished by Keenan et al. \shortcite{KeenanCII}.
Test calculations showed that our results seem to be in good agreement with
their values, although it is not possible to make an accurate statement of the
discrepancies, since they have published their results only in graphical form.

\subsection{The atom O$^0$}

The ground state of the O$^0$ atom is comprised of the 2s$^2$2p$^4$ $^3\mathrm{P}^e_{2,1,0}$
triplet levels. The energies of the fine-structure excited levels relatively to the ground state
are 158.265 cm$^{-1}$ and 226.977 cm$^{-1}$. The transition probabilities are
$A_{12}=8.865\ 10^{-5}\ \mathrm{s}^{-1}$,
$A_{02}=1.275\ 10^{-10}\ \mathrm{s}^{-1}$ and
$A_{01}=1.772\ 10^{-5}\ \mathrm{s}^{-1}$.

Our model atom includes the five lowest energy levels:
2s$^2$2p$^4$ $^3\mathrm{P}^e_{2,1,0}$,
2s$^2$2p$^4$ $^1\mathrm{D}^e_2$ and
2s$^2$2p$^4$ $^1\mathrm{S}^e_0$.
The energies were taken from
Moore \shortcite{MooreOI} and the transition probabilities from the Iron Project calculation
of Galav\'\i s, Mendoza and Zeippen \shortcite{AijCI}.

As the fine-structure levels of atomic oxygen are much more separated compared
to atomic and singly ionized carbon, the CMBR will not play a major role as
one can see from the excitation rates for the first excited level 
given in table \ref{table:Kij} (the excitation rates for the second
excited level are even lower).

\begin{figure}
\vbox{
\psfig{figure=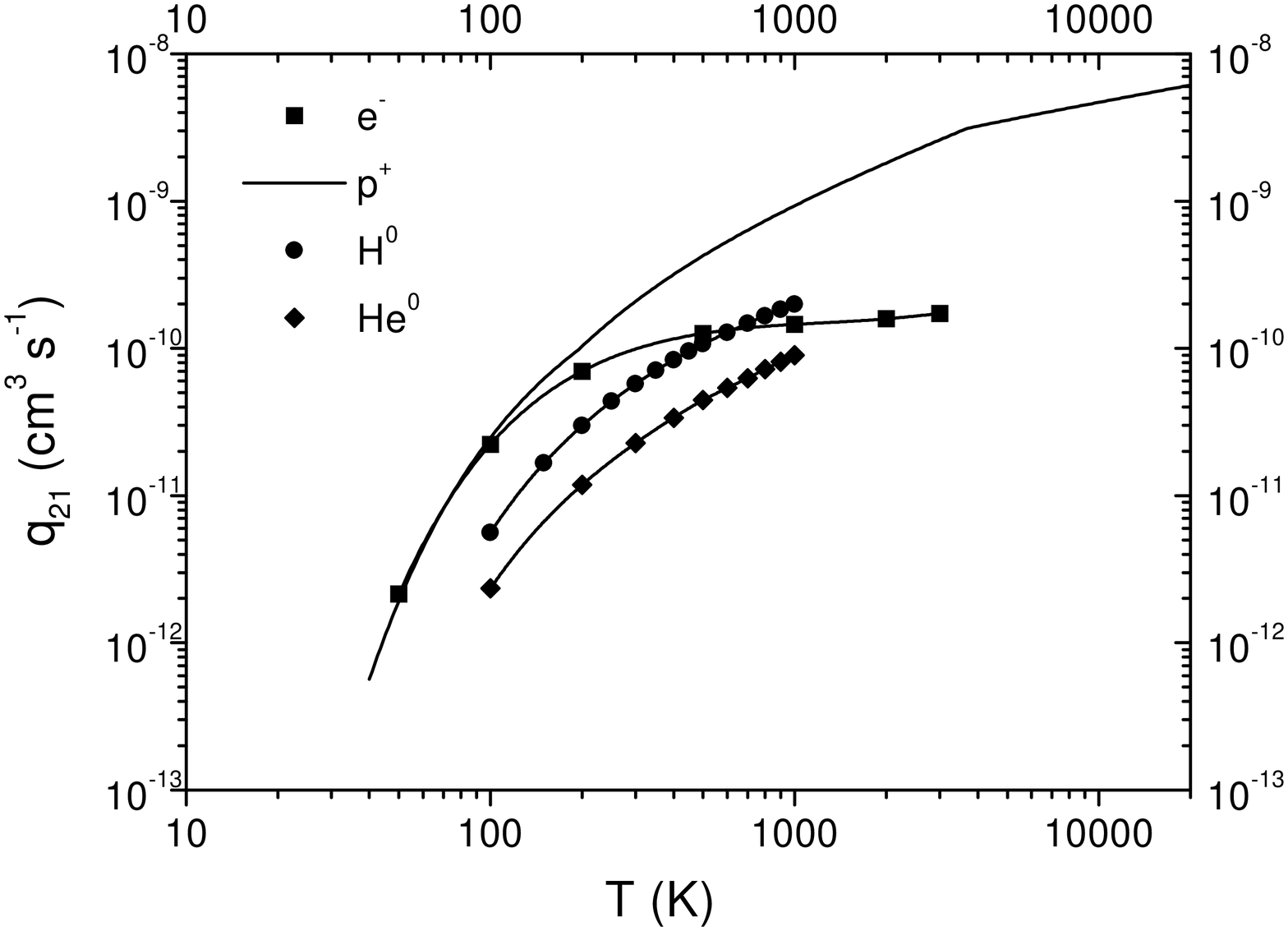,width=9.5cm}
\psfig{figure=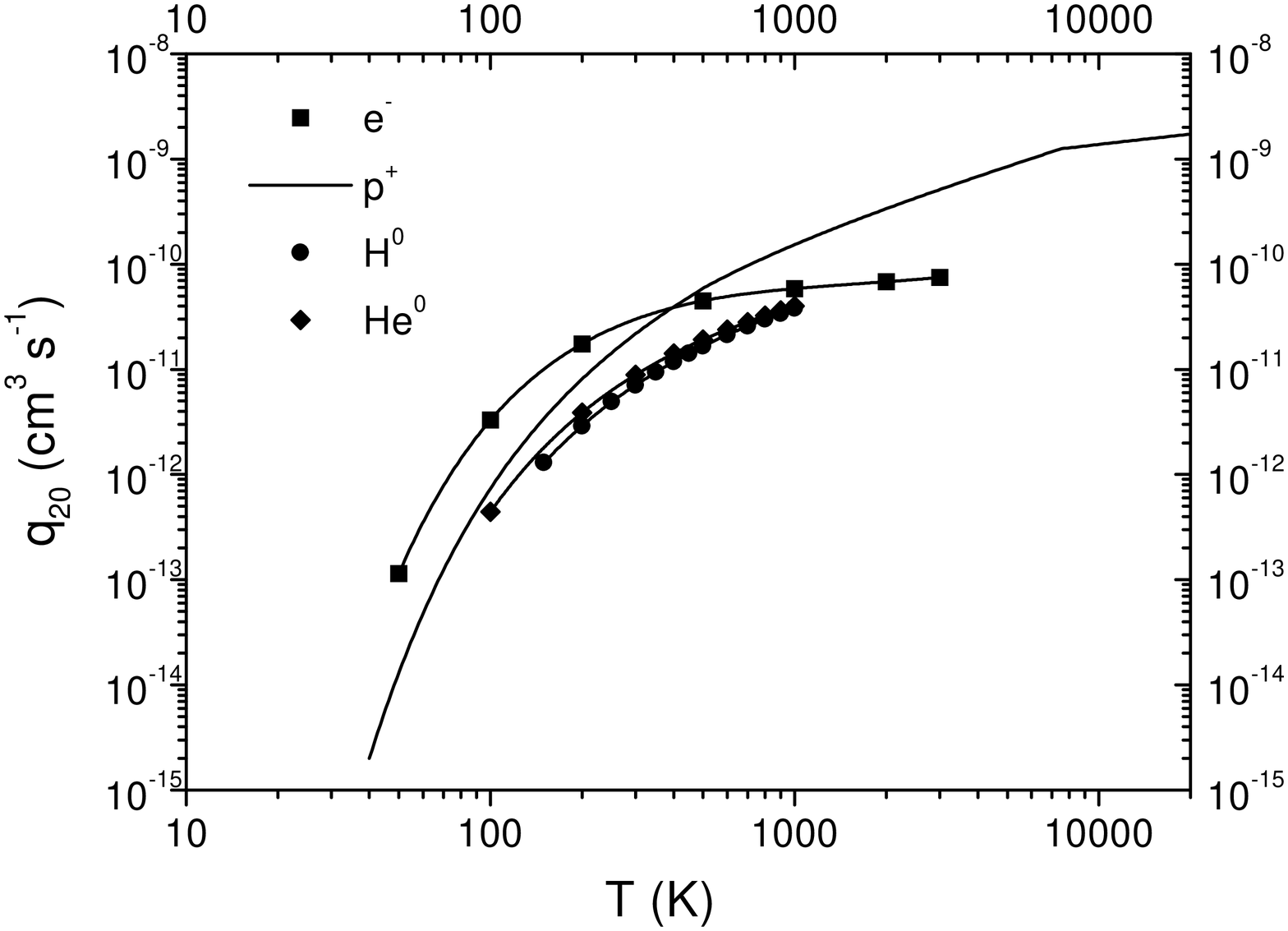,width=9.5cm}
\psfig{figure=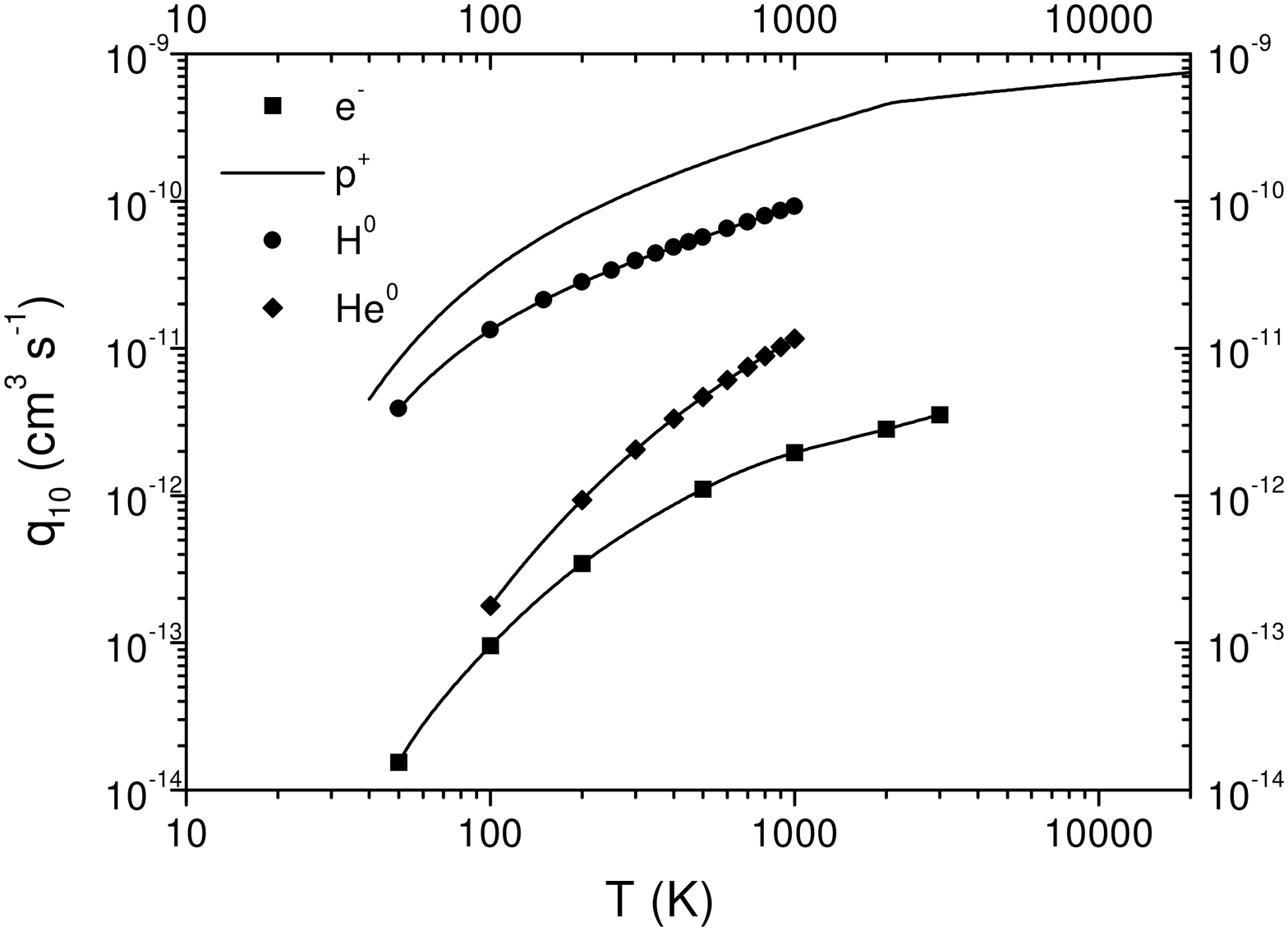,width=9.5cm}
}
\caption{Excitation rates $q_{JJ'}=q({^3P^e_J}\rightarrow{^3P^e_{J'}})$
of the O$^0$ fine-structure levels by collisions with various particles.
The points - taken from the literature cited in the text - are interpolated by
cubic splines.}
\label{figure:qijOI}
\end{figure}

The excited levels may be populated by collisions with particles present in the medium.
Fig. \ref{figure:qijOI} shows the collision rates for the fine-structure transitions
induced by collisions with various particles. The rates for collisional excitation
by electrons were taken from Bell, Berrington \& Thomas \shortcite{electrons_OI},
by neutral hydrogen from Launay \& Roueff \shortcite{q_H0_CI} and by neutral helium
from Monteiro \& Flower \shortcite{He_OI}.
For collisions with protons we have employed the analytic fits given by P\'equignot
\shortcite{Peq90,Peq96}.

For the sake of completeness, we have also considered collisional excitation of the upper
$^1\mathrm{D}^e_2$ and $^1\mathrm{S}^e_0$ levels.
We have taken the Maxwellian-averaged collision strengths for transitions
induced by electrons involving these levels from Berrington \& Burke \shortcite{BB81}.
The rate for the $^3\mathrm{P}^e$-$^1\mathrm{D}^e$ transition induced by neutral hydrogen
was taken from Federman \& Shipsey \shortcite{FS83}. The rates were transformed from
LS coupling to the individual fine-structure levels according to eq. (\ref{eq:LSindividual}).

After including 135 UV allowed transitions involving the ground
$^3\mathrm{P}^e$ levels and upper levels from the work of Verner et al.,
we obtained the indirect excitation rates by the radiation field of the
Galaxy: $\Gamma_{21}=3.9\ 10^{-11}\ \mathrm{s}^{-1}$
and $\Gamma_{20}=1.1\ 10^{-11}\ \mathrm{s}^{-1}$.

The relative populations of the ground $^3\mathrm{P}^e_J$ levels may be
significantly affected by charge exchange reactions with hydrogen
\cite{Peq90,Peq96}:

\begin{eqnarray}
& &\rmn{H}^+ +\, \rmn{O}^0({^3\rmn{P}^e_J}) \rightarrow
\rmn{H}^0({^2\rmn{S}^e_{1/2}}) +\, \rmn{O}^+({^4\rmn{S}^o_{3/2}}) \\
& &\rmn{H}^0({^2\rmn{S}^e_{1/2}}) +\, \rmn{O}^+({^4\rmn{S}^o_{3/2}})
\rightarrow
\rmn{H}^+ +\, \rmn{O}^0({^3\rmn{P}^e_{J'}}) \, . \nonumber
\end{eqnarray}

Consideration of this process would require a knowledge of the ionization
state of cloud, which lies beyond the scope of this paper.
Therefore, in our analysis we consider only the case of a primarily neutral
medium\footnote{This is the case for the DLA systems (section \ref{section:DLA}),
where \hbox{O\,{\sc i}} lines are commonly observed.}.

\begin{figure*}
\vbox{
\hbox{
\psfig{figure=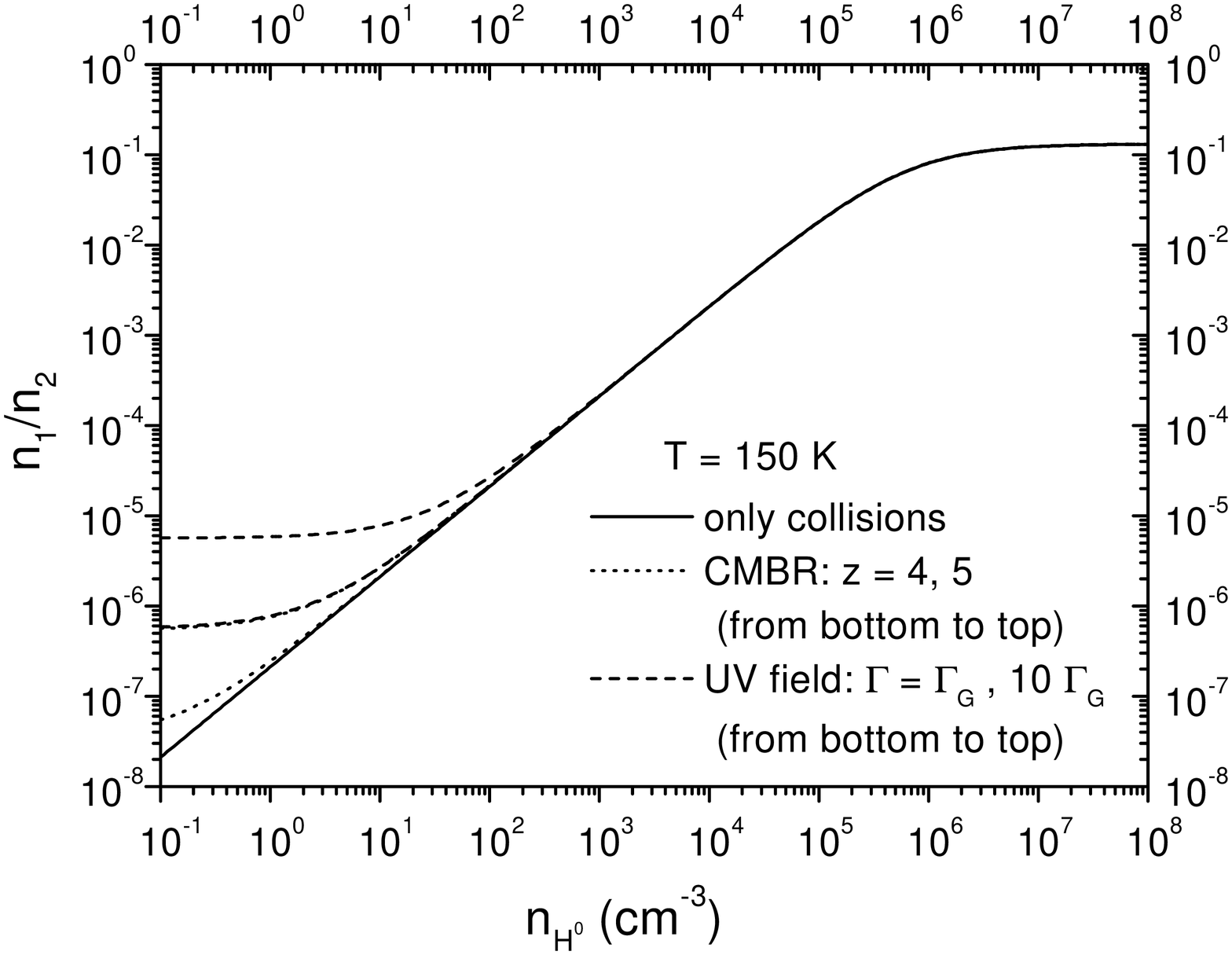,width=9.5cm}
\psfig{figure=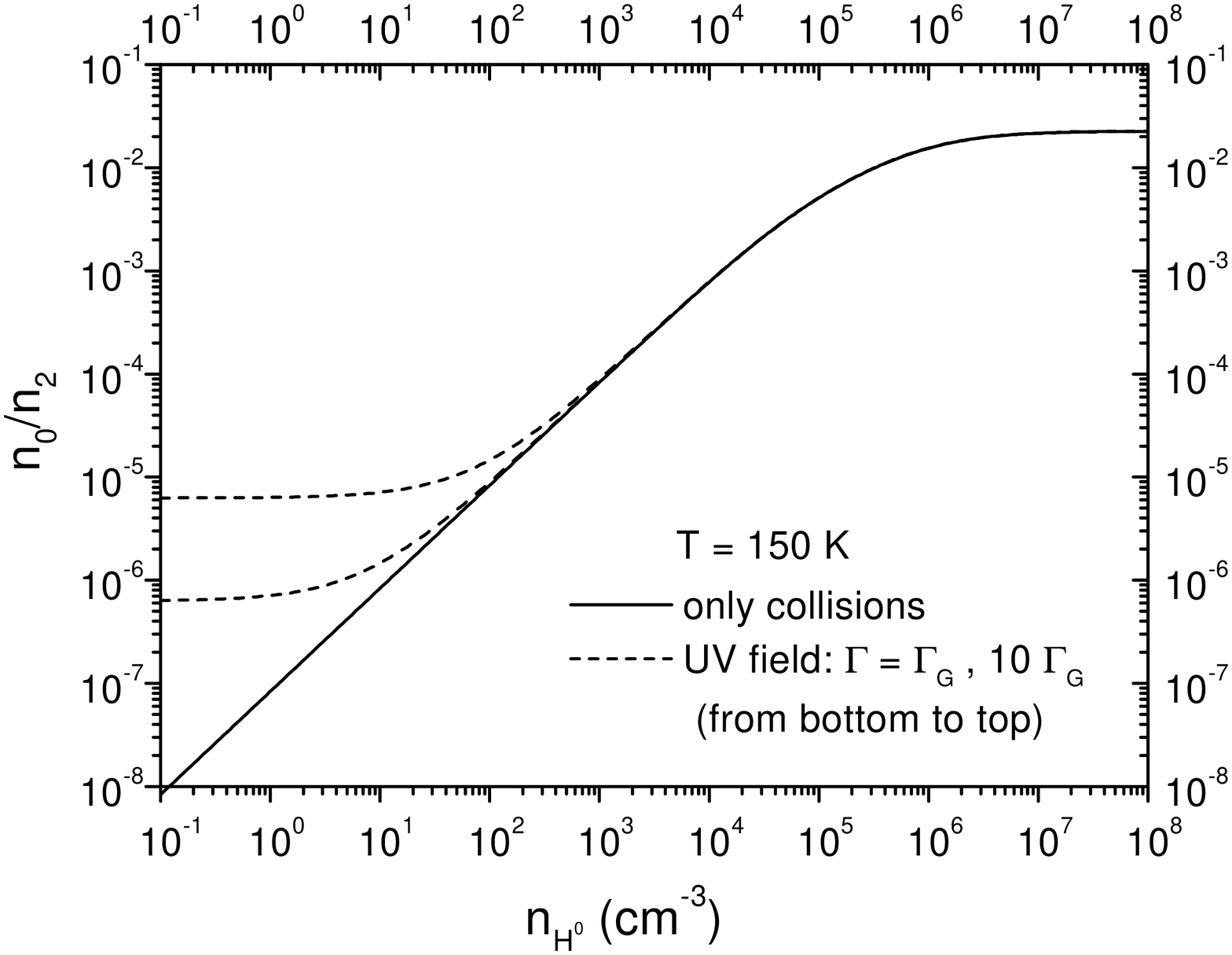,width=9.5cm}
}
\hbox{
\psfig{figure=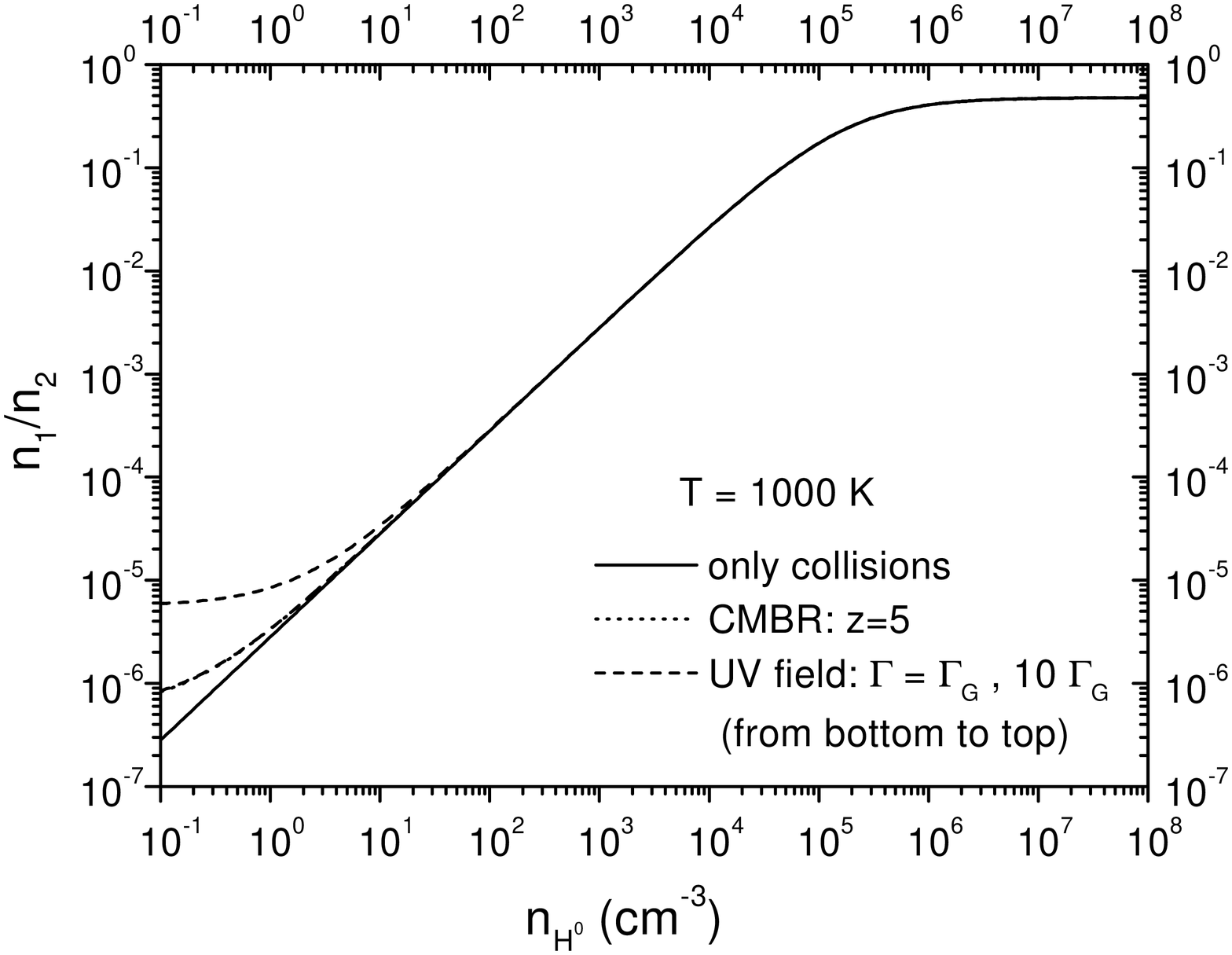,width=9.5cm}
\psfig{figure=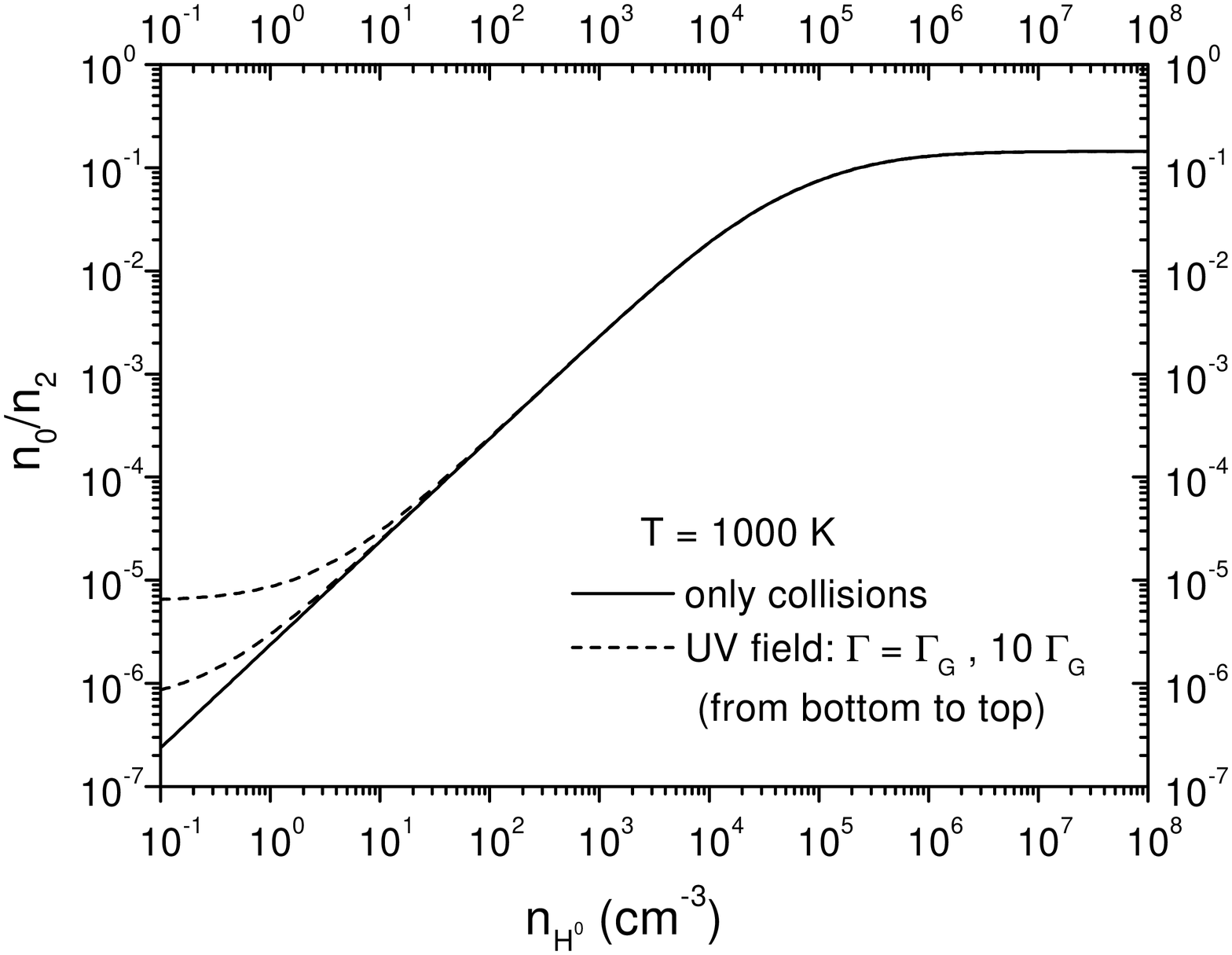,width=9.5cm}
}
}
\caption{Population ratios of the O$^0$ fine-structure levels relatively
to the ground state $n_J/n_2 = n({^3\mathrm{P}^e_J})/n({^3\mathrm{P}^e_2})$
calculated under various physical conditions.}
\label{figure:popOI}
\end{figure*}

In fig. \ref{figure:popOI} we plot the population ratios of the ground
O$^0$ fine-structure levels under various physical conditions.
We consider collisions by hydrogen and helium atoms, assuming a helium
abundance relative to hydrogen of 10 percent (by number).
Collisions by helium atoms increases
$n({^3\mathrm{P}^e_1})/n({^3\mathrm{P}^e_2})$ by only 5 percent and
$n({^3\mathrm{P}^e_0})/n({^3\mathrm{P}^e_2})$ by 10 percent (reducing to
zero close to LTE in the high density limit).
The curves for $n({^3\mathrm{P}^e_1})/n({^3\mathrm{P}^e_2})$ corresponding
to the inclusion of the effects of the CMBR at $z=5$ and the UV field of
the Galaxy are coincident because the relevant excitation rates are of the
same order $K^{z=5}_{21} \cong \Gamma^{\rmn{G}}_{21}$.

Our results are not directly comparable to the work of P\'equignot
\shortcite{Peq90,Peq96}, who made assumptions on the ionization state of
the gas. We point out, however, the importance of updating the electron
excitation rates employed in their work - taken from Berrington
\shortcite{Berrington88} - to the more recent calculations of Bell et al.,
since the later's results are substantially lower.

\subsection{The ion Si$^+$}

The ground state of the Si$^+$ ion consists of the
3s$^2$3p $^2\mathrm{P}^o_{\frac{1}{2},\frac{3}{2}}$
doublet levels. The energy of the fine-structure excited level relatively to the ground state
is  287.24 cm$^{-1}$, and the transition probability is
$A_{\frac{3}{2}\frac{1}{2}}=2.17\ 10^{-4}\ \mathrm{s}^{-1}$.

Our model ion includes the three lowest LS terms:
3s$^2$3p $^2\mathrm{P}^o$,
3s 3p$^2$ $^4\mathrm{P}^e$ and
3s 3p$^2$ $^2\mathrm{D}^e$,
making a total of 7 levels when the fine-structure is accounted for.
The energies were taken from Martin and Zalubas \shortcite{E_Si_II}.
The transition probabilities for the
$^2\mathrm{P}^o_{\frac{3}{2}}\rightarrow{^2\mathrm{P}^o_{\frac{1}{2}}}$
forbidden transition was taken from Nussbaumer \shortcite{Nuss77},
those for the $^4\mathrm{P}^e\rightarrow{^2\mathrm{P}^o}$
intercombination transitions from Calamai, Smith \& Bergeson \shortcite{CSB1993}
and those for the $^2\mathrm{D}^e\rightarrow{^2\mathrm{P}^o}$
allowed transitions from Nahar \shortcite{Nahar98}.

Since the fine-structure levels of Si$^+$ are too separated apart from eachother,
the CMBR will not be an important excitation mechanism. Even for extremely high
redshifts $z=5$, the excitation rate was found to be just
$K_{\frac{1}{2}\frac{3}{2}}=4.7\ 10^{-15}\ \mathrm{s}^{-1}$.

\begin{figure}
\psfig{figure=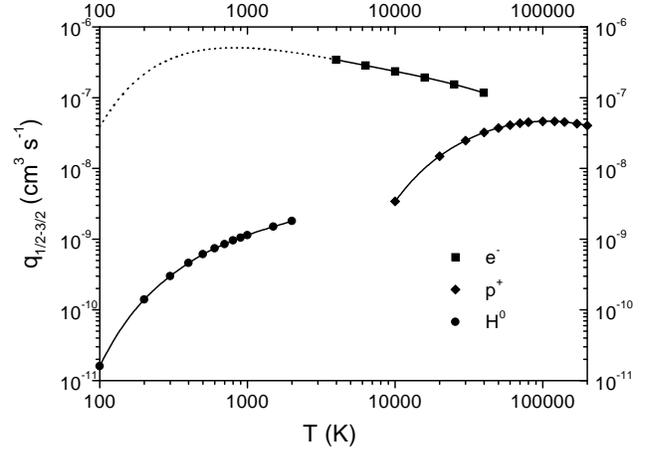,width=9.5cm}
\caption{Excitation rates $q_{\frac{1}{2}\frac{3}{2}}=
q({^2P^o_{\frac{1}{2}}}\rightarrow{^2P^o_{\frac{3}{2}}})$
of the Si$^+$ fine-structure level by collisions with various particles.
The points - taken from the literature cited in the text - are interpolated by
cubic splines. The dotted line indicates an extrapolation of the excitation rate
by electrons assuming a constant value for the corresponding Maxwellian-averaged
collision strength.}
\label{figure:qijSiII}
\end{figure}

Collisional processes considered are collisions by electrons \cite{electron_Si_II},
protons \cite{p_Si_II} and hydrogen atoms \cite{H0_Si_II}.
In fig. \ref{figure:qijSiII} we have plotted the excitation rates by collisions with
these particles.

Because the Maxwellian-averaged collision strength for the
$^2\mathrm{P}^o_{\frac{1}{2}}-{^2\mathrm{P}^o_{\frac{3}{2}}}$
transition induced by electrons varies by no more than
6 percent in the calculated interval - $3.6\leq \log T\leq 4.6$ - we also indicate
in fig. \ref{figure:qijSiII} what might be expected for the excitation rate down to
$T=100$ K if we assume a constant value for the collision strength.

In order to account for fluorescence, we consider 39 UV allowed transitions from the
work of Nahar \shortcite{Nahar98}. The indirect excitation rate by the UV field of the
Galaxy was found to be $\Gamma_{\frac{1}{2}\frac{3}{2}}=1.1\ 10^{-9}\ \mathrm{s}^{-1}$.

At sufficiently high temperatures the $^4\mathrm{P}^e$ and $^2\mathrm{D}^e$ upper levels
may be populated through collisions with hot electrons in the medium, and thereby
influence the population of the $^2\mathrm{P}^o$ ground levels.
To assess the relevance of this effect, we performed test calculations of the population
ratios of the $^2\mathrm{P}^o$ fine-structure levels comparing the results obtained
by the 2-level ion with those by the 7-level ion. Only collisions by electrons and spontaneous
decays were considered.
The test calculations revealed that the upper levels are not important for $T\leq 30000$ K,
when the discrepancies reach about only 6 percent.
Therefore, as for C$^+$, for temperatures lower than this, only two levels can be taken into
account. The system of statistical equilibrium equations (\ref{eq:sum}) then yields:

\begin{equation}
\frac{n_{\frac{3}{2}}}{n_{\frac{1}{2}}} =
\frac{Q_{\frac{1}{2}\frac{3}{2}}}{Q_{\frac{3}{2}\frac{1}{2}}}
\cong \frac{\Gamma_{\frac{1}{2}\frac{3}{2}}+
\sum_k n^k q^k_{\frac{1}{2}\frac{3}{2}}}
{A_{\frac{3}{2}\frac{1}{2}} + \sum_k n^k q^k_{\frac{3}{2}\frac{1}{2}}} \ .
\end{equation}

Excitation and de-excitation collisional rates are related by:

\begin{equation}
q_{\frac{3}{2}\frac{1}{2}} =
\frac{1}{2} q_{\frac{1}{2}\frac{3}{2}} \mathrm{e}^{\frac{413.27}{T}} \ ,
\end{equation}

with $T$ expressed in K.

\begin{figure}
\vbox{
\psfig{figure=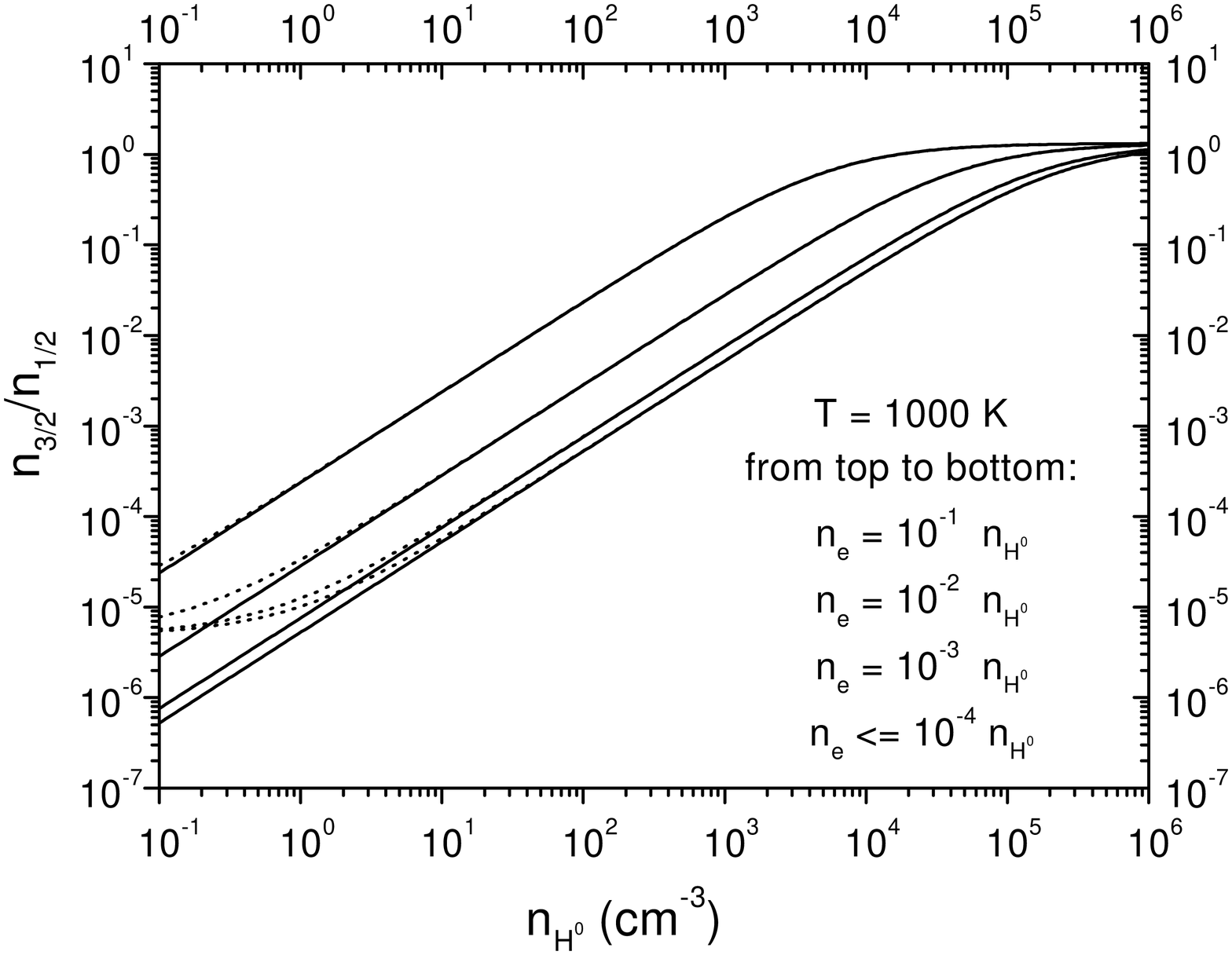,width=9.5cm}
\psfig{figure=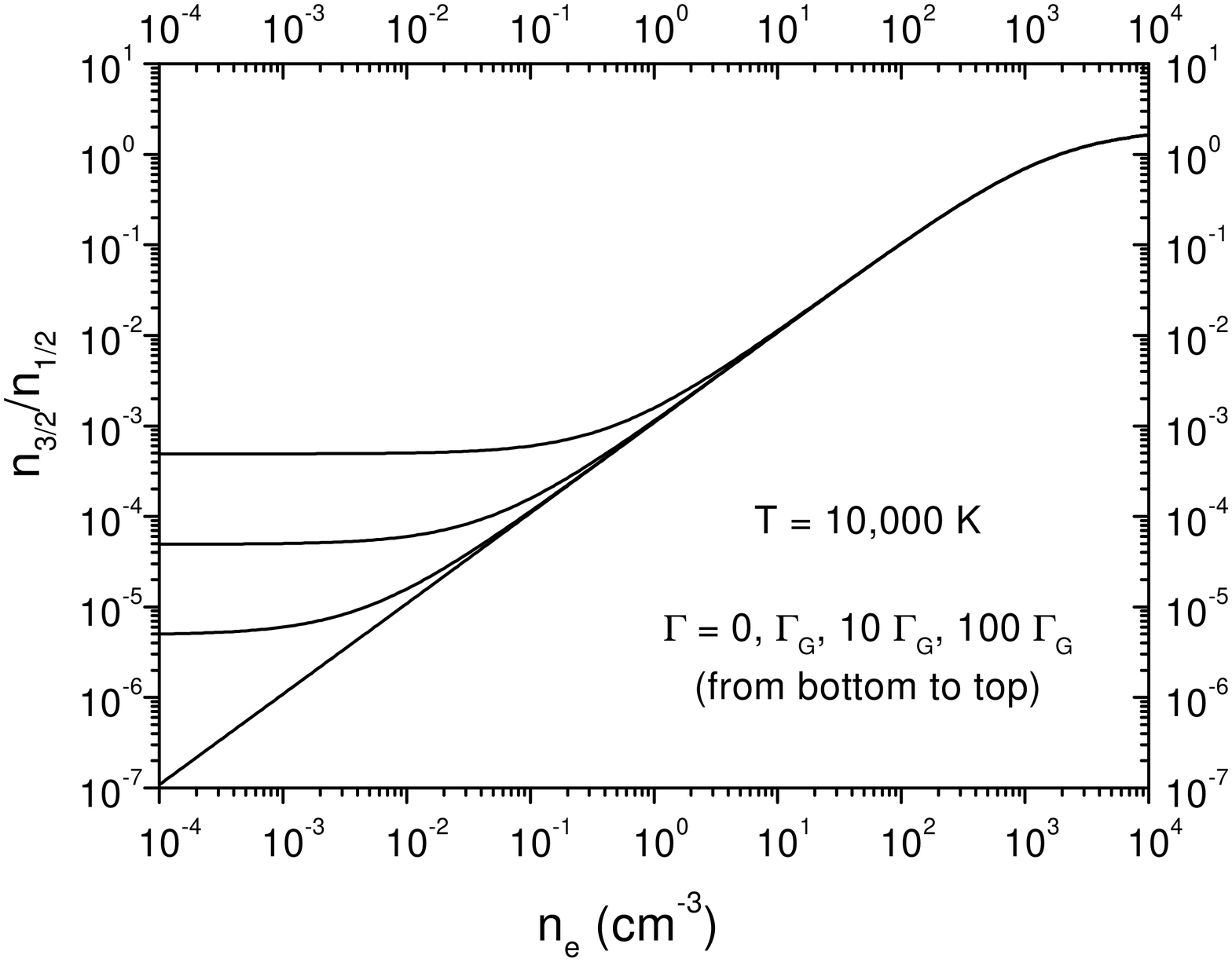,width=9.5cm}
}
\caption{Population ratio of the Si$^+$ fine-structure level relatively to the
ground state $n_{\frac{3}{2}}/n_{\frac{1}{2}} =
n({^2\mathrm{P}^o_{\frac{3}{2}}})/n({^2\mathrm{P}^o_{\frac{1}{2}}})$
calculated under various physical conditions. In the dotted curves in the upper
plot we have also added the contribution of fluorescence induced by the UV
field of the Galaxy.}
\label{figure:popSiII}
\end{figure}

In fig. \ref{figure:popSiII} we plot the population ratios of the fine-strutucture
levels of Si$^+$ under various physical conditions. As Si$^+$ may be the prevailing
ionization state in both \hbox{H\,{\sc i}} and \hbox{H\,{\sc ii}} regions,
we sample two cases of interest: a neutral medium
at $T=1000$ K, and an ionized medium at $T=10000$ K.

Previous calculations of the population ratios of the ground fine-structure levels
of Si$^+$ were performed by Keenan et al. \shortcite{pop_SiII}.
Although test calculations under the same physical conditions considered by the authors
appeared to reveal general agreement, it is difficult to quantify the discrepancies,
since they published their results in graphical form only.
We recommend the present calculations to the users, since they are based on more detailed
and accurate atomic data.

\subsection{The ion Fe$^+$}

The ground state of the Fe$^+$ ion is comprised of the
3d$^6$4s $^6\mathrm{D}^e_{\frac{9}{2},\frac{7}{2},\frac{5}{2},\frac{3}{2},\frac{1}{2}}$
sextet levels.
Comparing to the other atoms/ions previously studied, the ion Fe$^+$ has its fine-structure
levels very separated apart from eachother and the transition probabilities are
considerably higher.
For example, the first excitated level is placed 384.790 cm$^{-1}$ above the ground level,
and the corresponding transition probability is
$A_{\frac{7}{2}\frac{9}{2}}=2.13\ 10^{-3}\ \mathrm{s}^{-1}$.
Both factors will contribute to make the population ratios of the fine-structure levels
of the Fe$^+$ ion significantly low.

Our model ion includes the four lowest LS terms:
3d$^6$4s $^6\mathrm{D}^e$,
3d$^7$ $^4\mathrm{F}^e$,
3d$^6$4s $^4\mathrm{D}^e$ and
3d$^7$ $^4\mathrm{P}^e$,
making a total of sixteen levels when the fine-structure splitting is accounted for.
The energies were taken from Corliss \& Sugar \shortcite{E_FeII} and the transition
probabilities from the Iron Project calculation of Quinet, Le Dourneuf \& Zeippen
\shortcite{Aij_FeII}
\footnote{We note that there is a small misprint for two transitions listed in table 5 of
Quinet et al.'s paper. The authors did not add to the transition probabilites
a magnetic dipole contribution, so that the values should actually read:
a $^4\mathrm{F}^e_{\frac{7}{2}}$ - a $^4\mathrm{P}^e_{\frac{5}{2}}
=8.83\ 10^{-3}\ \mathrm{s}^{-1}$,
a $^4\mathrm{F}^e_{\frac{5}{2}}$ - a $^4\mathrm{P}^e_{\frac{5}{2}}
=1.68\ 10^{-3}\ \mathrm{s}^{-1}$, as it appears in their table 4.}.

Due to the high separation of the fine-structure levels, the CMBR will not be an
important excitation mechanism. For example, at $z=5$, the excitation rate to
the first excited level is just
$K_{\frac{9}{2}\frac{7}{2}}=3.5\ 10^{-18}\ \mathrm{s}^{-1}$.

\begin{figure}
\psfig{figure=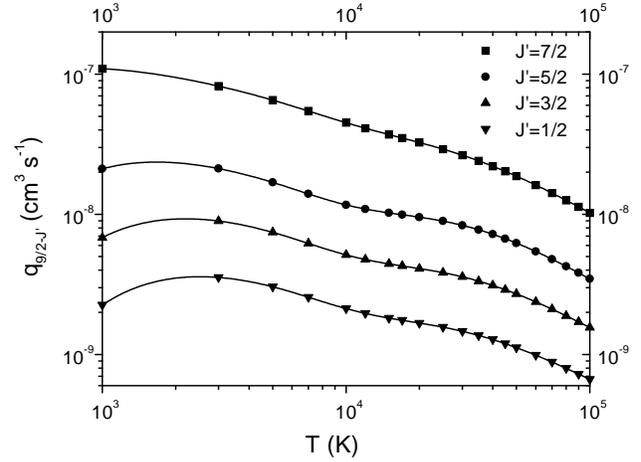,width=9.5cm}
\caption{Excitation rates $q_{\frac{9}{2}J'}=q({^6D^e_{\frac{9}{2}}}\rightarrow{^6D^e_{J'}})$
of the Fe$^+$ ground level to excited fine-structure levels by collisions with electrons.
The corresponding Maxwellian-averaged collision strengths were taken from the Iron Project
calculation of Zhang \& Pradhan \shortcite{q_e_FeII}, and were interpolated by
cubic splines.}
\label{figure:qijFeII}
\end{figure}

The only collisional process for which we could find detailed excitation rates calculated
in the literature were collisions by electrons. Fig. \ref{figure:qijFeII} shows
the excitation rates by collisions with electrons for the most important transitions
within the $^6\mathrm{D}^e$ ground term.
The corresponding Maxwellian-averaged collision strengths were taken from the Iron Project
calculation of Zhang \& Pradhan \shortcite{q_e_FeII}.

Nussbaumer \& Storey \shortcite{NussStorey} estimated the excitation rates by collisions
with protons to be less than 10 percent of the corresponding excitation rates by collisions
with electrons for temperatures as high as $T=15000$ K. However, as it is apparent from
fig. \ref{figure:qijCII} and fig. \ref{figure:qijSiII}, the excitation rates for
collisional processes involving positive ions and protons increase rapidly with temperature,
so that one should be cautious when negleting collisions by protons at extremely high
temperatures.

We also include 212 allowed transitions involving the $^6\mathrm{D}^e$ ground term levels
and upper levels from the Iron Project calculation of Nahar \shortcite{gam_FeII}.
The indirect excitation rates by fluorescence induced by the UV field of the
Galaxy were found to be:
$\Gamma_{\frac{9}{2}\frac{7}{2}}=7.0\ 10^{-10}\ \mathrm{s}^{-1}$,
$\Gamma_{\frac{9}{2}\frac{5}{2}}=1.3\ 10^{-10}\ \mathrm{s}^{-1}$ and
$\Gamma_{\frac{9}{2}\frac{3}{2}}=\Gamma_{\frac{9}{2}\frac{1}{2}}=0$
(the later rates are zero because those transitions are not
eletric dipole allowed).

Previous calculations of the fine-structure population ratios of Fe$^+$ levels were
performed by Keenan et al. \shortcite{pop_FeII}. They included only the two lowest
LS states in their model ion, 3d$^6$4s $^6\mathrm{D}^e$ and 3d$^7$ $^4\mathrm{F}^e$,
arguing that the next two LS states, 3d$^6$4s $^4\mathrm{D}^e$ and
3d$^7$ $^4\mathrm{P}^e$, do not affect significantly the population of the
$^6\mathrm{D}^e$ ground levels. However, test calculations showed that considering
the later LS terms increases the $^6\mathrm{D}^e$ ground level population ratios
by as much as 17 percent for $T=10000$ K.
The tests consisted of comparing the results of the
9-level and 16-level model ions taking only collisions by electrons 
(over various volume densities) and spontaneous decays into account.
Therefore one should use the 16-level model ion for $T>10000$ K.

In order to assess the relevance at high temperatures of even higher-lying levels in
the population ratios of the $^6\mathrm{D}^e$ ground levels, we have expanded our
16-level model ion to include the next two LS terms:
3d$^7$ $^2\mathrm{G}^e$ and
3d$^7$ $^2\mathrm{P}^e$, thereby increasing the total number of levels to 20.

Due to limitations of space, Quinet et al. list transition probabilities just for the
strongest ($A_{ij}>10^{-3}\ \mathrm{s}^{-1}$) transitions involving these levels.
Hence, for the sake of completeness, we decided to complement their work with the
weaker transitions from Garstang \shortcite{Aij_weak_FeII}.
The Maxwellian-averaged collision strengths for these transitions were taken from
the Iron Project calculation of Bautista \& Pradhan \shortcite{gam_2_FeII}.

Test calculations with the 20-level model ion revealed that the last two LS terms
do not affect at all the population ratios of the $^6\mathrm{D}^e$ ground levels
up to $T\cong 20000$ K (the highest temperature considered in Bautista \& Pradhan's
calculation).

\begin{figure}
\psfig{figure=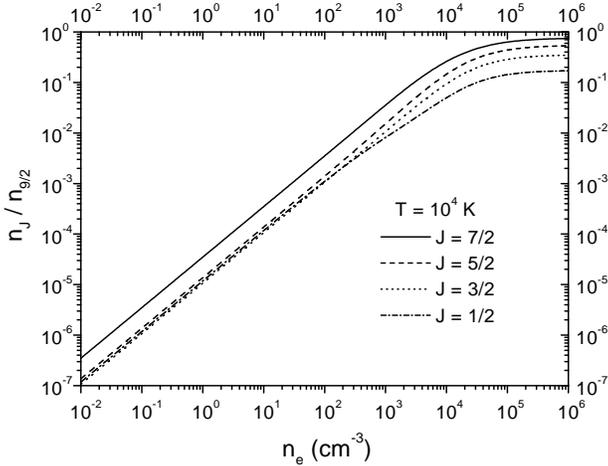,width=9.5cm}
\caption{Population ratio of the Fe$^+$ fine-structure levels relatively to the
ground state $n_J/n_{\frac{9}{2}} =
n({^6\mathrm{D}^e_J})/n({^6\mathrm{D}^e_{\frac{9}{2}}})$
as a function of electronic density.}
\label{figure:popFeII}
\end{figure}

Fig. \ref{figure:popFeII} shows the population ratios of the Fe$^+$ fine-structure levels
as a function of electronic density for $T=10000$ K.
We may note a slight inversion in the population
of levels ${^6\mathrm{D}^e_{\frac{3}{2}}}$ and ${^6\mathrm{D}^e_{\frac{1}{2}}}$
at lower densities.

Comparison of our results with those from the previous calculation of Keenan et al.
under the same physical conditions revealed that our values for the population ratios
of the fine-structure ground levels are a factor of 2-4 larger.
We believe this can be traced back to the Maxwellian-averaged collision strengths employed,
since the values from Zhang \& Pradhan are much higher than those obtained by Berrington
et al. \shortcite{old_gam_FeII}, quoted by Keenan et al. Since the Iron Project calculation
of Zhang \& Pradhan delineates the resonance structure of the collision strengths in more
detail, the results obtained in the calculation presented here should be more reliable.

\section{Physical Conditions}
\label{section:physicalconditions}

We now proceed to use our calculated atomic level population ratios to
study the physical conditions in QSO absorbers.

Table \ref{obsdata} shows our sample of absorption line systems for which
there are column density ratios of fine-structure lines reported in the
recent literature.

\begin{table*}
\centering
\begin{minipage}{181mm}
\caption{Observational data on the column density ratios of fine-structure
lines in QSO absorbers retrieved from the literature.}
\label{obsdata}
\begin{tabular}{@{}llllllllll@{}}
\#  & QSO         & $z_{\rmn{em}}$ & $z_{\rmn{abs}}$ & $\log$\,N(\hbox{H\,{\sc i}}) & ion               & $N^*/N$ $^{\rmn{a}}$                 & $T_{\rmn{exc}}$ $^{\rmn{a}}$ & $T_{\rmn{cmbr}}$ $^{\rmn{b}}$ & reference \\
\hline
1   & PKS 1756+23 & 1.721          & 1.6748          & $>\,20.3$                    & \hbox{C\,{\sc i}} & $<\,1.2\,10^{-1}$ $^{\rmn{c}}$           & $<\,7.4$             & 7.289            & Roth \& Bauer \shortcite{RB} \\
2a  & Q1331+17    & 2.084 $^{\rmn{d}}$          & 1.77638         & 21.2 $^{\rmn{d}}$                       & \hbox{C\,{\sc i}}  & $(3.1\,\pm\,0.3)\,10^{-1}$   & $10.4\,\pm\,0.5$    & 7.566            & Songaila et al. \shortcite{Songb} \\
2b  & Q1331+17    & 2.084 $^{\rmn{d}}$          & 1.77654         & 21.2 $^{\rmn{d}}$                      & \hbox{C\,{\sc i}}  & $(1.3\,\pm\,0.4)\,10^{-1}$   & $7.4\,\pm\,0.8$     & 7.566            & Songaila et al. \shortcite{Songb} \\
3   & Q0013-00    & 2.0835 $^{\rmn{e}}$         & 1.9731          & 20.7 $^{\rmn{e}}$                      & \hbox{C\,{\sc i}}  & $(4.0\,\pm\,0.8)\,10^{-1}$   & $11.7\,\pm\,1.1$    & 8.102            & Ge, Bechtold \& Black \shortcite{Ge97} \\
3   & Q0013-00    & 2.0835 $^{\rmn{e}}$         & 1.9731          & 20.7 $^{\rmn{e}}$                      & \hbox{C\,{\sc ii}} & $(7.0\,\pm\,3.2)\,10^{-3}$   & $16.2\,\pm\,1.3$    & 8.102            & Ge, Bechtold \& Black \shortcite{Ge97} \\
4   & Q0149+33    & 2.43           & 2.140           & 20.5 $^{\rmn{f}}$                      & \hbox{C\,{\sc ii}} & $<\,9.6\,10^{-3}$             & $<\,17.1$            & 8.557            & Prochaska \& Wolfe \shortcite{PW99} \\
5   & Q1946+76    & 2.994          & 2.8443          & 20.27                     & \hbox{C\,{\sc ii}} & $<\,2.1\,10^{-2}$             & $<\,20.0$            & 10.476           & Lu et al. \shortcite{Lu96b} \\
6   & Q0636+68    & 3.174 $^{\rmn{g}}$           & 2.9034          & 17.7 $^{\rmn{h}}$                      & \hbox{C\,{\sc ii}} & $<\,6.7\,10^{-3}$ $^{\rmn{i}}$            & $<\,16.0$            & 10.637           & Songaila et al. \shortcite{Songa} \\
7   & Q0347-38    & 3.23           & 3.025           & 20.7 $^{\rmn{d}}$                      & \hbox{C\,{\sc ii}} & $<\,2.8\,10^{-2}$             & $<\,21.4$            & 10.968           & Prochaska \& Wolfe \shortcite{PW99} \\
8   & Q2212-16    & 3.992          & 3.6617          & 20.2                      & \hbox{C\,{\sc ii}} & $<\,2.3\,10^{-2}$             & $<\,20.5$            & 12.703           & Lu et al. \shortcite{Lu96b} \\
9   & Q2237-06    & 4.559          & 4.0803          & 20.5                      & \hbox{C\,{\sc ii}} & $<\,4.4\,10^{-3}$             & $<\,14.9$            & 13.844           & Lu et al. \shortcite{Lu96b} \\
10  & BRI 1202-07 & 4.7            & 4.3829          & 20.6                      & \hbox{C\,{\sc ii}} & $<\,1.2\,10^{-2}$             & $<\,18.0$            & 14.668           & Lu et al. \shortcite{Lu96a}
\end{tabular}

\medskip
$^{\rmn{a}}$ Errors are $1\sigma$ CL, while upper limits are $2\sigma$ CL.

$^{\rmn{b}}$ Assuming the temperature-redshift relation predicted by the
standard model.

$^{\rmn{c}}$ $2\sigma$ upper limit on $N^*$ obtained by private communication
with the author.

$^{\rmn{d}}$ Pettini et al. \shortcite{P94}.

$^{\rmn{e}}$ Ge \& Bechtold \shortcite{GB97}.

$^{\rmn{f}}$ Wolfe et al. \shortcite{W93}.

$^{\rmn{g}}$ Sargent, Steidel \& Boksenberg \shortcite{Sargent89}.

$^{\rmn{h}}$ Derived from the optical depth of the LL discontinuity:
$\tau_{\rmn{LL}}=3.5$ \cite{Sargent89}.

$^{\rmn{i}}$ Given the strong saturation of the ground fine-structure
line, we adopt $N>1.5\,10^{14}\,\rmn{cm}^{-2}$ instead of the profile
fitting value $N=4.6\,10^{14}\,\rmn{cm}^{-2}$ preferred by Songaila et al.
\shortcite{Songa}.
\end{minipage}
\end{table*}

The sample includes DLA systems
($\log\rmn{N}(\hbox{H\,{\sc i}})>20.3$), and only one LL
system at $z_{\rmn{abs}}=2.9034$.

We have not included any associated system,
since their close proximity to the QSO could make them
susceptible to the influence of the background radiation source, therefore
requiring a case by case analysis that lies beyond the scope of this paper.
The fine-structure lines are, however, a valuable tool to infer the physical
conditions in such systems. In particular, the knowledge of the ionization
state of the systems coupled with the information on the volumentric density
afforded by the fine-structure lines allows one to place limits on the
distance between the absorber and the QSO, giving a clue to infer whether
they correspond to intervening clouds or to material ejected from the QSO
\cite{TWW79,Morris,TLS96,SP2000}.

So far, all the fine-structure lines observed belong to either C$^0$ or
C$^+$. Owing to its low ionization fraction (since its ionization potential
is lower than that of hydrogen), atomic carbon is very seldom detected.
The three systems listed in table \ref{obsdata} correspond to all
of the presently known \hbox{C\,{\sc i}} systems, apart from the system
observed towards the BL Lac object 0215+015 \cite{Bladesa,Bladesb}.

As we gathered observational data from the literature, we rejected any line
falling within the Ly-$\alpha$ forest region of the spectrum.
Prochaska \shortcite{P99} observed the \hbox{C\,{\sc ii}$^*$} 1335
fine-structure transition in a LL system at $z_{\rmn{abs}}=2.652$ towards
Q2231-00. However, since this transition falls within the Ly-$\alpha$
forest in this object and therefore may have been subject to significant
contamination, his claimed value on the column density
N(C\,{\sc ii}$^*$) should be regarded at most as an upper limit to the true
value. For the same reason we disregarded the DLA system at
$z_{\rmn{abs}}=3.054$ towards Q0000-26 observed by Giardino \& Favata
\shortcite{GF2000}. Although the authors quoted their value for
N(C\,{\sc ii}$^*$) as an upper limit, we argue that in principle significant
contamination could also be taking place on the ground fine-structure line,
thereby also affecting N(C\,{\sc ii}) and driving the ratio $N^*/N$ in the
opposite sense.

Unfortunately, the ground \hbox{C\,{\sc ii}} 1334 line is often heavily
saturated; to circunvent this problem
there have been many alternative approaches to derive the N(C\,{\sc ii})
column density by other indirect methods.
Prochaska \shortcite{P99} used the ratio of N(C\,{\sc ii})/N(Fe\,{\sc ii})
in a velocity region where the ground \hbox{C\,{\sc ii}} line 
was not saturated
to derive the corresponding value at the component where the
\hbox{C\,{\sc ii}$^*$} line was detected. Outram, Chaffee \& Carswell
\shortcite{OCC} assumed a carbon abundance relative to iron [C/Fe]$>$-0.3
to obtain a tighter lower limit on the N(C\,{\sc ii}) column density
in a DLA system at $z_{\rmn{abs}}=2.62$ towards GB1759+75.
In our sample we have included only direct measurements on the column
densities.

In sections \ref{section:DLA}-\ref{section:LL} below,
we will separately study the DLA and LL systems in our sample.
Again, as a working hypothesis we shall assume
the temperature-redshift relation as predicted by the standard model.
The validity of this relation is discussed in section \ref{section:CMBR}.

\subsection{DLA systems}
\label{section:DLA}

DLA systems have very high neutral hydrogen column densities
($\log\rmn{N}(\hbox{H\,{\sc i}})>20.3$).
This makes them effectively shielded from the ionizing radiation, causing
their contents to be essentially neutral material
\cite{Viegas95}.

We use the fine-structure lines column density ratios observed in the
DLA systems listed in table \ref{obsdata} to set upper limits to their
neutral hydrogen volume densities $n_{\rmn{H}^0}$ and to the
intensities of the UV radiation field present. Given the high neutral hydrogen
column density, probably all of the hydrogen ionizing radiation will be
absorbed, leaving very few photons with energies greater than 1 Ryd.
The spectral shape of the UV radiation field will then be similar to
the one found in our own galaxy, and we therefore assume the UV
radiation field of Gondhalekar et al. multiplied by a constant factor
$f_{\rmn{G}}$.

Table \ref{DLAconditions} shows the upper limits to $n_{\rmn{H}^0}$ and
$f_{\rmn{G}}$ for the DLA systems in our sample. They represent firm upper
limits to the true values, because the single excitation mechanism
considered to obtain the upper limit - i.e., collisions by neutral
hydrogen atoms to obtain $n_{\rmn{H}^0}$ and fluorescence to obtain
$f_{\rmn{G}}$ - may not be the dominating one and also because for most
systems the population ratios were just upper limits.

\begin{table*}
\centering
\begin{minipage}{134mm}
\caption{Physical conditions in DLA systems. The quoted values for
$f_{\rmn{G}}$ and $n_{\rmn{H}^0}$ are upper limits, whereas those for
$l$ and $M$ are lower limits. The CL is $2\sigma$. The second figure
next to each entry corresponds to the inclusion of the CMBR (if only one
value appears, it is not altered to the last digit displayed).}
\label{DLAconditions}
\begin{tabular}{@{}cccccccc@{}}
    &               & \multicolumn{3}{c}{$T=100$ K} & \multicolumn{3}{c}{$T=1000$ K} \\
\#  & $f_{\rmn{G}}$ & $n_{\rmn{H}^0}$ [cm$^{-3}$] & $l$ [pc] & $M$ [$M_{\sun}$] & $n_{\rmn{H}^0}$ [cm$^{-3}$] & $l$ [pc] & $M$ [$M_{\sun}$] \\
\hline
1  & 16 / 0.79 & 16 / 0.79 & 4.0 / 81    & 26 / 10600  & 6.9 / 0.34 & 9.4 / 192 & 142 / 58800 \\
2a & 42 / 25   & 43 / 26   & 11 / 19     & 1540 / 4290 & 18 / 11    & 27 / 46   & 8910 / 24900 \\
2b & 16 / -    & 16 / -    & 30 / -      & 10700 / -   & 6.9 / -    & 70 / -    & 59600 / - \\
4  & 236 / 235 & 35 / 34   & 2.8         & 19          & 12         & 7.8       & 146 / 148 \\
5  & 518 / 510 & 77 / 75   & 0.79 / 0.80 & 0.93 / 0.96 & 27         & 2.2       & 7.3 / 7.5 \\
7  & 703 / 691 & 105 / 103 & 1.6         & 10          & 37 / 36    & 4.4       & 77 / 79 \\
8  & 582 / 544 & 86 / 81   & 0.60 / 0.64 & 0.45 / 0.52 & 31 / 29    & 1.7 / 1.8 & 3.6 / 4.1 \\
9  & 107 / 39  & 16 / 5.7  & 6.2 / 17    & 93 / 692    & 5.7 / 2.1  & 17 / 47   & 711 / 5320 \\
10 & 307 / 209 & 45 / 31   & 2.9 / 4.2   & 26 / 57     & 16 / 11    & 8.0 / 12  & 205 / 445
\end{tabular}
\end{minipage}
\end{table*}

Because the collisional excitation rate is temperature dependent, so will
be the derived upper limits on $n_{\rmn{H}^0}$; we assume two values
of kinetic temperature characteristic of \hbox{H\,{\sc i}} regions:
$T=100$ K and $T=1000$ K.

If excitation by the CMBR is taken into account the upper limits become
tighter (lower), as indicated by the second figure next to each entry in
table \ref{DLAconditions} (if only one value appears, it remains unchanged
to the last significant digit displayed). Accounting for the CMBR affects
considerably the results for the \hbox{C\,{\sc i}} systems (objects 1 and 2),
since it is an important excitation mechanism for C$^0$ as mentioned earlier
in section \ref{C0}. Note the striking difference between both values for
object 1, which has an excitation temperature very close to the predicted
CMBR temperature. We can not consider the CMBR for object 2b, since the
excitation temperature is slightly lower than the CMBR temperature.
As for the remaining \hbox{C\,{\sc ii}} systems the result is changed
significantly only for the $z>4$ regions (objects 9 and 10), when the
CMBR starts to play a significant role at the neutral hydrogen densities
involved (cf. fig. \ref{figure:popCII}, top).

Collisions with molecular hydrogen are not likely to be relevant in our
analysis, since the molecular fraction usually seen in DLA systems is
exceedingly small: $f(\rmn{H}_2)\equiv2N(\rmn{H}_2)/N(\rmn{H})<2\,10^{-4}$,
reaching as low as $f(\rmn{H}_2)=4\,10^{-8}$ in the $z_{\rmn{abs}}=3.3901$
DLA system towards Q0000-26 \cite{Lev2000}.
Two exceptions are object 3 in our sample \cite{GB97} and the
$z_{\rmn{abs}}=2.34$ DLA system towards Q1232+08 \cite{Ge2000}, with
$f(\rmn{H}_2)=0.22$ and 0.07, respectively. In any case that would imply
$n_{\rmn{H}_2}q^{\rmn{H}_2}_{ij}<<n_{\rmn{H}^0}q^{\rmn{H}^0}_{ij}$,
as typically $q^{\rmn{H}_2}_{ij}<q^{\rmn{H}^0}_{ij}$
(cf. figs. \ref{figure:qijCI} and \ref{figure:qijCII}).

From table \ref{DLAconditions} we see that the ratio of fine-structure
lines observed in DLA systems constrain their neutral hydrogen densities
to be lower than tens of cm$^{-3}$ (or a few cm$^{-3}$ in the best cases),
and the UV radiation field to be lower than two orders of magnitude the
radiation field present in our galaxy (or one order of magnitude in the
best cases).

Naturally, we could also have placed upper limits to the electron density
$n_e$. The upper limits on $n_e$ derived from \hbox{C\,{\sc ii}} lines
would be about two orders of magnitude lower than the corresponding upper
limits on $n_{\rmn{H}^0}$ listed in table \ref{DLAconditions}, i.e., in
the approximate inverse ratio of the corresponding collision rates
$q^e_{\frac{1}{2}\frac{3}{2}}/q^{\rmn{H}^0}_{\frac{1}{2}\frac{3}{2}}
\approx10^2$ (fig. \ref{figure:qijCII}). For C$^0$ this ratio is no more
than 10 in the relevant temperature region (fig. \ref{figure:qijCI}).
As DLA systems are comprised of mostly neutral material, the free electrons
will come mainly from neutral atoms which have an ionization potential
lower than that of hydrogen, such as C$^0$, an whose (solar) elemental
abundance relative to hydrogen is of order $10^{-4}$.
Therefore, we would expect beforehand $n_e\approx10^{-4}n_{\rmn{H}^0}$,
and the fine-structure lines would not provide a meaninful constraint
on the electron density.
The electron/neutral hydrogen density ratio may be even lower if we
consider that DLA systems may exhibit abundances as low as two orders
of magnitude below solar (e.g. Pettini et al. 1994).

We can also estimate the characteristic sizes and total masses of the
absorbing clouds responsible for the DLA systems; these will be given
by:

\begin{eqnarray}
l & = & \frac{\rmn{N}(\rmn{H})}{n_{\rmn{H}}} \\
M & = & m_p n_{\rmn{H}}\,l^3 = m_p \rmn{N(H)}\,l^2\, , \nonumber
\end{eqnarray}

where $m_p$ is the proton's mass and N(H) and $n_{\rmn{H}}$ are the
total hydrogen column and volume densities, respectively.

For DLA systems we can make the replacement $\rmn{N(H)}\cong \rmn{N(H\,{\sc i})}$
and $n_{\rmn{H}}\cong n_{\rmn{H}^0}$
\footnote{Some authors use the \hbox{C\,{\sc ii}} fine-structure lines
to constrain $n_e$ and set $n_{\rmn{H}}\cong n_e$; from the discussion
in the preceding paragraph we note that this underestimates
$n_{\rmn{H}}$ by two orders of magnitude.}
; hence we can use our upper limits on
$n_{\rmn{H}^0}$ to set lower limits to the characteristic sizes and
total masses of the intervening clouds.
We see from table \ref{DLAconditions} that our constraints imply
characteristic sizes larger than a few pc (tens of pc in the best cases) and
lower limits for the total masses that vary from $10^0$ to $10^5$ solar masses.

In deriving the cloud sizes and masses above, we have implicitly assumed
that most of the hydrogen column density is in the same component where
the fine-structure lines could be measured. Altough it would be much more
difficult to detect the excited fine-structure line in the velocity
component with the lowest associated hydrogen column density, we can not
rule out the possibility that this is compensated by a higher metalicity
and intensity of local excitation mechanisms.

\begin{figure}
\psfig{figure=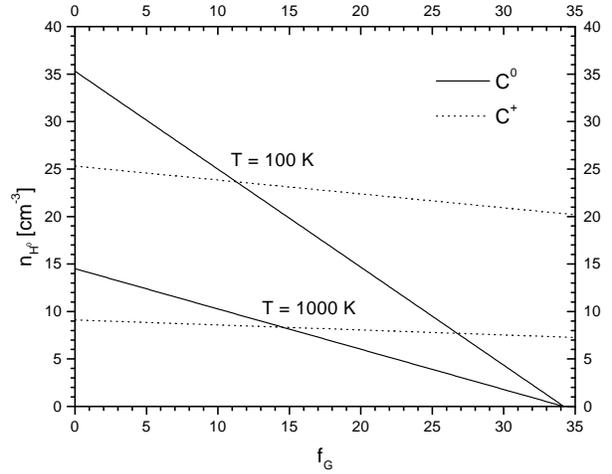,width=9.5cm}
\caption{Physical conditions in object 3 in our sample
(the CMBR is included).}
\label{figure:obj3}
\end{figure}

We now focus our attention to object 3 in our sample, which exhibits both
\hbox{C\,{\sc i}} and \hbox{C\,{\sc ii}} fine-structure lines.
In fig. \ref{figure:obj3} we derive the neutral hydrogen volume density as
a function of the intensity of the UV radiation field based upon the
column density ratios of \hbox{C\,{\sc i}} and \hbox{C\,{\sc ii}} lines
(and for the two values of kinetic temperature considered before;
the CMBR is included).
If we assume that C$^0$ and C$^+$ are located within the same ionization
region in the cloud, then the physical conditions will be described by
the intersection of both curves.
For $T=100$ K we have $n_{\rmn{H}^0}=24$ cm$^{-3}$ and $f_G=11$,
whereas for $T=1000$ K we have $n_{\rmn{H}^0}=14$ cm$^{-3}$ and $f_G=8$.
Therefore, regardless of the kinetic temperature adopted, the UV field
present must be one order of magnitude more intense than in our galaxy.
This contrasts with object 1, where the observed \hbox{C\,{\sc i}} lines
constrain the UV field to be lower than in our galaxy.
In the detailed photoionization model constructed by Ge et al. the physical
conditions prevailing in most regions of the cloud are:
$T=100$ K, $n_{\rmn{H}^0}=21.0\pm9.6$ cm$^{-3}$, $f_G=17.0$ and
$n_e=5.0\,10^{-4}n_{\rmn{H}^0}$.
Assuming their value for $n_{\rmn{H}^0}$ we have $l=7.7\pm3.6$ pc and
$M=240\pm160\,M_{\sun}$.

\subsection{LL systems}
\label{section:LL}

The LL systems differ considerably from the DLA systems studied before
for beeing significantly ionized. The source of the ionizing radiation
in these systems is usually assumed to be the UV extragalactic background,
as the integrated radiation field of all QSOs attenuated by the
intergalactic medium \cite{HM96}.

Some authors claim for a local origin to the source of ionization \cite{VF95}.
In particular, the puzzling observations of column densities of C, N and O
ions in various ionization states as well as of \hbox{He\,{\sc i}} in LL
systems towards Q1700+64 \cite{VR93,RV93} cannot be simultaneously explained
by photoionization models based on the UV background as the source of
ionization. These systems are successfully interpreted by the {\em hot
halo model} \cite{VF95}; in this model the LL systems are identified as cold
condensations embedded in a hot halo formed during the early stages of
galaxy evolution, which acts as the source of ionization.

Photoionization models typically constrain the ionization parameter
$U\equiv\frac{\Phi_{\rmn{H}}}{n_{\rmn{H}}c}$ ($\Phi_{\rmn{H}}$ is the
total photon flux of hydrogen ionizing radiation); the fine-structure
lines might be used to independently constrain the volumetric density
and test the hypothesis of a given radiation field as beeing the source of
ionization. If the density determined from the fine-structure lines,
together with the ionization parameter determined from ionization models
based on the UV background imply a higher intensity for the ionizing
radiation field than expected,
that would be a strong evidence for a local origin to the true ionization
source.

For the only LL system in our sample (object 6 in table \ref{obsdata}),
we derive a $2\sigma$ upper limit to the electronic density of
$n_e<0.15$ cm$^{-3}$, assuming a kinetic temperature
\hbox{$T=10^4$ K} characteristic of photoionized regions.
We have included the minor contribution from the CMBR and collisions by
protons (assuming a fully ionized medium $n_p=n_e$), although they affect
the result only at the 10 percent level.
Fluorescence plays a negligible role. In table \ref{table:hothalo} we show
the indirect excitation rates of C$^+$ fine structure levels for the
hot halo models considered by Viegas \& Fria\c ca; in any case we have
$\Gamma_{\frac{1}{2}\frac{3}{2}}<<\,n_eq^e_{\frac{1}{2}\frac{3}{2}}$.
The indirect excitation rate induced by the UV background turns out to be
even lower; we have adopted
the revised calculation of Madau, Haardt \& Rees
\shortcite{MHR99} to obtain an indirect excitation rate at the observed
redshift of $\Gamma_{\frac{1}{2}\frac{3}{2}}=3.1\,10^{-12}$ s$^{-1}$.

\begin{table}
\caption{Indirect excitation rates of C$^+$ fine-structure levels by the
radiation fields predicted by hot halo models. Each model, taken from
Viegas \& Fria\c ca, corresponds to a given age and distance from the
center of the forming galaxy.}
\label{table:hothalo}
\begin{tabular}{@{}ccc}
$t$ (Gyr) & $r$ (kpc) & $\Gamma_{\frac{1}{2}\frac{3}{2}}$ (s$^{-1}$) \\
\hline
0.206  & 30  & 6.8\ 10$^{-10}$ \\
0.206  & 100 & 6.2\ 10$^{-11}$ \\
0.3644 & 30  & 1.3\ 10$^{-11}$ \\
0.3644 & 100 & 1.2\ 10$^{-12}$
\end{tabular}
\end{table}

\subsection{The CMBR temperature-redshift relation}
\label{section:CMBR}

The CMBR constitutes one of the cornerstones of the hot Big Bang model,
which makes three basic quantitative predictions on its properties:
\begin{enumerate}
\item it is isotropic and homogeneous;
\item it has a black-body spectrum,
\item it cools as the universe expands according to the relation
$T=T_0 (1+z)$.
\end{enumerate}

Over the past decade, the advent of the {\it COBE} satellite has allowed
the confirmation of the isotropy \cite{Smoot92} and black-body spectral
shape \cite{Mather94} to unprecedented precision, giving a present day
temperature of $T_0=2.725\pm 0.001$ K ($1\sigma$ error) as determined
from the {\it FIRAS} instrument \cite{Mather99,Smoot}.

Any direct means of measuring the CMBR temperature can just provide us with
the value of its current temperature, forcing us to resort to other indirect
methods to test the temperature-redshift relation predicted by the standard
model. The best alternative is to use atomic and molecular transitions seen
in the spectra of QSO absorbers \cite{Meyer94}. It is worth noting that the
observation of CN absorption lines from diffuse interstellar clouds towards
bright stars in the Galaxy yielded $T_0=2.729^{+0.023}_{-0.031}$ K,
in excellent agreement with the {\it COBE FIRAS} result
\cite{Roth92,Roth93,Roth95}.

Unfortunately, molecular transitions are not commonly seen in the spectra
of QSO absorbers. Apart from H$_2$, so far molecules have been
identified in just four absorption systems \cite{WC94,WC95,WC96a,WC96b}.
Surprisingly, in one of them \cite{WC96b} the rotational transitions
from several molecules indicated an excitation temperature
$T_{\rmn{exc}}=4\pm 2$ K ($3\sigma$ error), lower than the expected CMBR
temperature $T=5.14$ K predicted at the observed redshift.
The low excitation temperature in this object is, nevertheless, due to
the effect of a microlensing event (Combes 2000, private comm.).
Molecular absorption systems are often gravitational lenses, since the
impact parameter to the foreground galaxy must be close to zero in order
to allow the detection of molecules. Hence, we believe that atomic lines
are better suited to study the temperature of the CMBR at high redshifts.

We can use the population ratios of the fine-structure levels for the
absorption systems collected in table \ref{obsdata} to constrain the
temperature of the CMBR at their redshifts. For each observed ion the
excitation temperature will be given by

\begin{eqnarray}
T_{\rmn{exc}} & = & \frac{23.60}{\ln(3/(n^*/n))}\qquad \rmn{for\,\,C}^0 \\
T_{\rmn{exc}} & = & \frac{91.25}{\ln(2/(n^*/n))}\qquad \rmn{for\,\,C}^+ \,, \nonumber
\end{eqnarray}

with temperatures given in K.

\begin{figure}
\psfig{figure=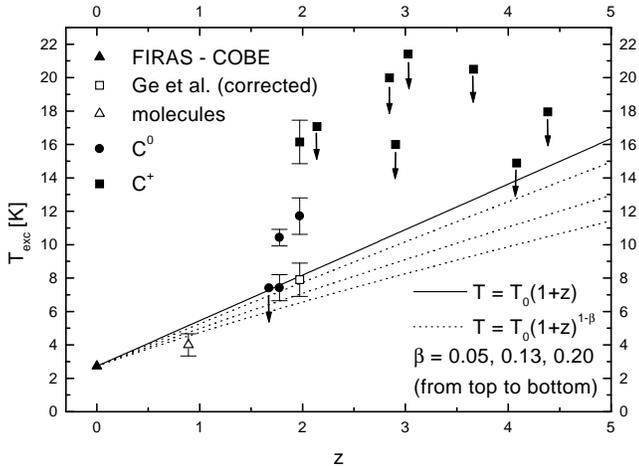,width=9.5cm}
\caption{Excitation temperatures derived from fine-structure absorption
lines. The solid line is the temperature of the CMBR according to the
temperature-redshift relation given by the standard model, while
alternative models with photon creation predict a lower temperature (dotted
lines). Error bars are $1\sigma$ CL, whereas upper limits are $2\sigma$ CL.}
\label{figure:Tcmbr}
\end{figure}

The excitation temperatures so obtained represent firm upper limits to
the temperature of the CMBR, because local excitation mechanisms may also
contribute significantly to populate the excited levels. In fig.
\ref{figure:Tcmbr} we plot the excitation temperatures along with the
expected temperature of the CMBR according to the standard model prediction.
For most systems, either the signal to noise ratio of the spectrum was not
high enough to detect the excited fine structure line, or the ground
\hbox{C\,{\sc ii}} line was strongly saturated. Therefore for these systems
the excitation temperature itself is also an upper limit, and this is
indicated in fig. \ref{figure:Tcmbr} by a downward arrow. The point
labelled "molecules" corresponds to the puzzling observation of Wiklind \&
Combes \shortcite{WC96b} discussed above.

Phillips \shortcite{Phil94} supports a closed steady-state model that predicts
considerably lower temperatures to the CMBR compared to the standard model;
e.g. for the $z=2.9$ \hbox{C\,{\sc ii}} system observed by Songaila et al.
\shortcite{Songa}: $T_{\rmn{obs}}=6.55$ K.

Alternative models in which photon creation takes place as the Universe
expands predict a more general temperature-redshift relation \cite{LSV2000}:

\begin{equation}
T=T_0(1+z)^{1-\beta}\,,
\label{Tbeta}
\end{equation}

where $\beta$ is a parameter to be adjusted from the observations, within the
range $0\leq\beta\leq 1$. Equation (\ref{Tbeta}) therefore gives
temperatures lower than predicted by the standard model
(fig. \ref{figure:Tcmbr}).

It has been stated that any scenario that does not preserve the number of
photons would introduce large distortions in the black-body spectrum of
the CMBR \cite{Steigman}. However, it was shown that for the class of models
that follow the temperature law (\ref{Tbeta}) the Planckian spectral shape
is not destroyed as the Universe evolves \cite{Lima1,Lima2}. In these models
the total entropy of Universe increases with time, but the entropy per
particle remains constant.

Big Bang nucleosynthesis arguments, however, severely limit the value of
the free parameter to $\beta<0.13$ \cite{BS97}.

Inspection of fig. \ref{figure:Tcmbr} reveals that current measurements do
not require any extra ingredients to the standard model, since the totality
of the points lie above the linear temperature law. However, a conclusive
statement could only be made after correcting for local excitation
mechanisms, in order to convert the excitation temperature upper limits to
the actual temperature of the CMBR. Altough some points in fig.
\ref{figure:Tcmbr} appear to be dangerously close to the standard model
prediction, it could be that additional local excitation mechanisms are
negligible compared to excitation by the CMBR in these systems.
In that case, the excitation temperature would provide a
direct measure of the CMBR temperature.

For object 3 in our sample, Ge et al. \shortcite{Ge97} constructed a
detailed photoionization model to account for the local excitation
mechanisms. They obtained \hbox{$T=7.9\pm1.0$ K}, whereas the standard model
prediction is \hbox{$T=8.102\pm0.003$ K} (fig. \ref{figure:Tcmbr}).

If the temperature law given by the standard model turns out to be incorrect,
it would pose a serious source of difficulty, since not even the presence
of a cosmological constant would alter the predicted temperature-redshift
relation \cite{LSV2000}. On the other hand, if it is confirmed by the
observations, that would add another success to its list of triumphs with
a bonus: since each absorbing region is located at a different site of the
Universe, we could also assess its {\em homogeneity}.

\section{Conclusions}
\label{section:conclusions}

We have presented new theoretical calculations of population ratios of
the ground fine-structure levels of C$^0$, C$^+$, O$^0$, Si$^+$ and Fe$^+$.
The literature was searched for the most recent and reliable atomic data
available to date. Various possible excitation mechanisms are taken into
account. We encourage the user to make use of the available Fortran
code to obtain accurate predictions in his/her applications, rather then
rely on approximate practical formulas that are valid only for a limited
range of physical conditions.

We have retrieved from the literature observational data on the column
density ratios derived from the fine-structure lines, and confronted them
with our theoretical calculations to infer the physical conditions
prevailing in DLA and LL systems.
Currently only \hbox{C\,{\sc i}} and \hbox{C\,{\sc ii}} fine-structure
lines have been observed; future detection of lines originating from less
excited atoms/ions such as O$^0$, Si$^+$ and Fe$^+$ (which also have
resonant lines redward the Ly-$\alpha$ forest) might aid to better
constrain the physical conditions.

For the DLA systems, the neutral hydrogen volumetric density is lower than
tens of cm$^{-3}$ (a few cm$^{-3}$ in the best cases) and the UV radiation
field is less intense than two orders of magnitude the UV field of the
Galaxy (one order of magnitude in the best cases).
Their characteristic sizes are higher than a few pc (tens of pc in the best
cases) and lower limits for their total masses vary from $10^0$ to $10^5$ solar masses.

For the only LL system in our sample, we derived $n_e<0.15$
cm$^{-3}$. As more observations become available, it may be possible to
use the information contained in the fine-structure lines to help determine
the nature of the source of ionization of these systems.

The fine-structure lines in QSO absorbers also provide a method to test the
temperature-redshift relation for the CMBR predicted by the standard model.
Current observations do not contradict the linear temperature law, although
a conclusive statement could only be made after accounting for local
excitation mechanisms in order to correct the excitation temperatures
to the actual temperature of the CMBR.
That would require a knowledge of the ionization state of the cloud, after
appropriate modeling by a photoionization code.

A substantial improvement from the theoretical standpoint could be achieved
by analysing together the ionization state of the cloud and the excitation
of the fine-structure levels, by coupling our code - {\sc popratio} - to
a larger photoionization code.
Presently, all studies based on the excitation of the fine-structure
levels were carried out separately from the photoionization modeling,
considering average physical conditions throughout the entire cloud
(e.g. Ge et al., Giardino \& Favata).

By carrying out both analyses simultaneously, we could further refine
our models and eliminate the need for all assumptions we have maden in
our analyses of
the fine-structure lines. The intensities of the excitation mechanisms could
vary across the cloud, and it would no longer be necessary to stick to the
simplistic case of a perfect homogeneous cloud, that we implicitly assumed
when we wrote down eq. (\ref{eq:ratio}). Moreover, we could generalize
our statistical equilibrium equations (\ref{eq:sum}) to include terms
which require the knowledge of the ionization state of the cloud, such
as recombination and charge exchange.
It would also be possible to account for optical depth effects, allowing
us to drop the assumption that all transitions considered are optically
thin.

On the observational side, there is also clearly a need for better
measurements, since for the great majority of the systems only upper
limits to the column density ratios are available.
As the future generations of more powerful telescopes equipped with high
resolution spectrographs continue to push the detection limits to even
weaker lines, more information could be available by observing atoms/ions
other than C$^0$ and C$^+$.

\section*{Acknowledgments}

We would like to thank S. Nahar, M. Bautista and F. Haardt for providing us
electronic versions of their data and K. Roth and F. Combes for helpful
information.
AIS acknowledges finantial support by the Brazilian agency FAPESP, under contract 
No. 99/05203-8. This work is partially supported by CNPq (304077/77-1) and
PRONEX/FINEP (41.96.0908.00).

\bsp

\label{lastpage}

\end{document}